\documentclass[a4paper,11pt]{article}
\usepackage{jheppub} % for details on the use of the package, please see the JINST-author-manual
\usepackage{soul}
\usepackage{tocloft}
\usepackage{lineno}
\usepackage{gensymb}
\usepackage{caption}
\usepackage{subcaption}
\usepackage{cancel}
\usepackage{slashed}
\pdfoutput=1 % if your are submitting a pdflatex (i.e. if you have
\usepackage{color,graphicx,slashed,multirow,hyperref,amssymb,amsmath,cleveref,multirow}
\usepackage{braket}
\usepackage{graphicx}
\usepackage[utf8]{inputenc}
\usepackage[section]{placeins}
\usepackage{xcolor,colortbl}
\usepackage{subcaption}
\usepackage{cleveref}
\usepackage[shortlabels]{enumitem}
\AtBeginDocument{\usepackage{booktabs}}
\usepackage{feynmp-auto}
\usepackage[normalem]{ulem}

\title{\boldmath Fermionic Dark Matter and Neutrino Masses in a Minimal $U(1)_{\rm PQ}$ Inverse Seesaw Framework}

\author[a]{Nakorn Thongyoi,}
\author[b]{Patipan Uttayarat,}
\author[a]{and Chakrit Pongkitivanichkul}
\affiliation[a]{Khon Kaen Particle Physics and Cosmology Theory Group (KKPaCT),\\
Department of Physics, Faculty of Science, Khon Kaen University,\\
123 Mitraphap Rd, Khon Kaen 40002, Thailand}
\affiliation[b]{Theoretical High-Energy Physics and Astrophysics Research Unit (ThEPA), Department of Physics, Srinakharinwirot University, 114 Sukhumvit 23 Rd.,\\ Wattana, Bangkok
10110, Thailand}

\emailAdd{nakorn.thongyoi@gmail.com}
\emailAdd{patipan@g.swu.ac.th}
\emailAdd{chakpo@kku.ac.th}

\abstract{
We propose a minimal extension of the Standard Model (SM) where a single global $U(1)_{\rm PQ}$ symmetry simultaneously generates the scales for the inverse seesaw mechanism and stabilizes a fermionic dark matter (DM) candidate. In this framework, the spontaneous breaking of the symmetry naturally yields a pseudo-Nambu-Goldstone boson---an axion-like particle (ALP)---which serves as the inevitable mediator between the dark and visible sectors. This setup generates light neutrino masses at the TeV scale without requiring inaccessible energy scales, while predicting a strong correlation between the effective Majorana mass and the CP-violating phase, enhancing its testability in future experiments. We demonstrate that the model satisfies stringent constraints from neutrino oscillations, DM relic density, and the recent muon $g-2$ measurements, with the relic density governed primarily by DM annihilation into SM fermions via the ALP portal. This work offers a unified, structurally consistent framework that addresses the neutrino mass and dark matter puzzles, providing distinct signatures for particle physics and cosmology.
}

\begin{document}
\maketitle
\flushbottom

\section{Introduction}
\label{sec:Intro}

The Standard Model (SM) has been a cornerstone in understanding elementary particles and their interactions. Despite its successes, certain phenomena remain unexplained, notably the non-zero masses of neutrinos and the existence of dark matter (DM).  
The SM’s failure to accommodate these observations has driven significant efforts to extend it, leading to a range of proposed mechanisms and models.

One of the most profound issues with the SM is its prediction of massless neutrinos. However, experimental evidence from Super-Kamiokande~\cite{Super-Kamiokande:1998kpq}, SNO~\cite{SNO:2002tuh}, KamLAND~\cite{KamLAND:2002uet}, Daya Bay~\cite{DayaBay:2024hrv}, RENO~\cite{RENO:2012mkc}, CHOOZ~\cite{CHOOZ:1999hei}, K2K~\cite{K2K:2002icj}, T2K~\cite{T2K:2011ypd}, and LSND~\cite{LSND:2001aii} confirms that neutrinos undergo flavor oscillations, and hence are massive. Traditional seesaw mechanisms have been proposed to explain the smallness of neutrino masses by introducing heavy right-handed Majorana neutrinos, which generate light neutrino masses suppressed by the heavy Majorana scale~\cite{Weinberg:1979sa,Mohapatra:1979ia,Schechter:1980gr,King:2003jb,Mohapatra:1986bd,Dev:2012sg}, commonly referred to as the seesaw scale. However, the traditional seesaw mechanism usually requires a seesaw scale around $M\sim10^{12}$ GeV, which greatly reduces the possibility to probe the model experimentally. On the other hand, the inverse seesaw model~\cite{King:2013eh,Deppisch:2015qwa,Gavela:2009cd,Ibarra:2010xw,Deppisch:2004fa,Dev:2009aw,Das:2012ze,Bonilla:2023egs,Bharadwaj:2024crt} offers a more testable neutrino mass generation mechanism. In this scenario, sterile neutrinos with small, yet technically natural, lepton number-violation (LNV) couplings are introduced. These small LNV couplings help to lower the seesaw scale down to within reach of current colliders. This approach has gained traction as a realistic and testable framework, with studies exploring its implications on lepton flavor violation (LFV) and collider phenomenology~\cite{Langacker:1980js,Davidson:2008bu,Altarelli:2010gt,Abada:2014vea, Deppisch:2015qwa,Alekhin:2015byh,Hisano:1995cp,Kuno:1999jp,Beacham:2019nyx}.

On the other hand, the existence of DM has been firmly established through astrophysical observations, such as galaxy rotation curves~\cite{ParticleDataGroup:2024cfk,Bertone:2004pz} and cosmic microwave background measurements~\cite{Planck:2018vyg,WMAP:2003elm,COBE:1992syq}. Various extensions of the SM have been proposed to accommodate DM, see Refs.~\cite{Feng:2010gw,Arcadi:2017kky,Bertone:2018krk,Cirelli:2024ssz,Bozorgnia:2024pwk,Balazs:2024uyj} for recent reviews. One of the most studied scenarios is the Higgs portal DM~\cite{Patt:2006fw}, thanks to its simplicity and predictive nature.  In this framework, the Higgs boson serves as a mediator between the DM and the SM sectors. Phenomenology of the Higgs portal DM has been studied in Refs.~\cite{Djouadi:2011aa,Bishara:2015cha,Arcadi:2019lka}. Due to its minimal structure, the Higgs portal model cannot accommodate neutrino masses without a further extension. Moreover, the necessity of imposing a global symmetry in order to solely stabilize DM may be considered an unsatisfactory feature of the model.

Naturally, one can extend this by introducing a global $U(1)_{\rm PQ}$ to stabilize DM. Due to non-perturbative effects, $U(1)_{\rm PQ}$ is expected to be broken into a shift symmetry which give rise to an existence of the axion-like particle (ALP)~\cite{Dine:1981rt, Peccei:1977hh,Weinberg:1977ma}. ALP can then act as a mediator between the DM and SM sectors which offer unique signatures in indirect detection experiment, cosmic microwave background distortions, and low energy precision measurements, which can differentiate them from Higgs portal models~\cite{Nomura:2008ru,Freytsis:2009ct,Bauer:2022rwf,Zhevlakov:2022vio}. The introduction of another scale in the model can be beneficial for the neutrino mass generation mechanism as well~\cite{Carvajal:2015dxa}.

Another historically interesting observable is the muon anomalous magnetic moment, \( g-2 \), whose precise measurements at Fermilab~\cite{Muong-2:2021ojo,Muong-2:2023cdq} and previously at Brookhaven~\cite{Muong-2:2006rrc} have long appeared to deviate from earlier SM predictions~\cite{Aoyama:2020ynm}. This tension, which once stood at nearly \( 5\sigma \), motivated significant theoretical and experimental efforts. However, a recent comprehensive update by the Muon \( g-2 \) Theory Initiative~\cite{Aliberti:2025beg}, incorporating improved lattice QCD evaluations and addressing tensions in \( e^+e^- \) data, has led to a revised SM prediction that is in agreement with the experimental average within uncertainties~\cite{Muong-2:2025xyk}. While the anomaly is no longer statistically significant, the muon magnetic moment remains a sensitive probe for new physics.

In this work, we build upon these developments by proposing a minimal SM extension that combines the inverse seesaw mechanism for neutrino mass generation and the ALP portal DM. 
Our model draws from the previous study that explored Higgs portals in the DM-neutrino mass model~\cite{Pongkitivanichkul:2019cvm}, aiming instead to utilize the ALP portal~\cite{Fitzpatrick:2023xks,Armando:2023zwz,Ghosh:2023tyz,Gutierrez:2021gol,DiMeglio:2021iec,Gola:2021abm,Deniverville:2020rbv,Kirpichnikov:2020tcf,deNiverville:2019xsx,deNiverville:2018hrc,Kaneta:2017wfh}, which offers new dynamics and observables. However, we demonstrate that by imposing the $U(1)_{\rm PQ}\times \mathbb{Z}_2$ symmetry, we can simultaneously generate the mass scales required for the inverse seesaw mechanism and stabilize a fermionic Dark Matter candidate. As a direct consequence, the spontaneous breaking of the $U(1)_{\rm PQ}$ symmetry yields a pseudo-Nambu-Goldstone boson (the ALP) whose natural couplings to the Standard Model allow it to serve as the mediator between the dark and visible sectors. In addition, the remaining $\mathbb{Z}_2$ symmetry stabilises the DM. In this framework, the ALP is not an ad hoc addition, but rather an intrinsic component of the symmetry breaking mechanism that makes the neutrino sector realistic. %\textcolor{red}{CP: We probably need to stress this fact in the model section again and of course in the conclusion.}
%\textcolor{blue}{The DM sector is extended by introducing $U(1)_{\rm PQ}$ to stabilize DM and the ALP coming from the chiral symmetry breaking act as a mediator.}

%Our model draws from the previous study that explored Higgs portals in the DM-neutrino mass model~\cite{Pongkitivanichkul:2019cvm}, aiming instead to utilize the ALP portal~\cite{Fitzpatrick:2023xks,Armando:2023zwz,Ghosh:2023tyz,Gutierrez:2021gol,DiMeglio:2021iec,Gola:2021abm,Deniverville:2020rbv,Kirpichnikov:2020tcf,deNiverville:2019xsx,deNiverville:2018hrc,Kaneta:2017wfh}, which offers new dynamics and observables. In this work, we also apply the recent constraint on muon $g-2$~\cite{Aliberti:2025beg} to the ALP portal model, which place a non-trivial constraint on the model parameter space. This approach not only addresses the neutrino mass and DM issues within a unified framework but also proposes testable signatures that could guide future experimental searches.

For the complete analysis, we calculate the relic density of DM, the indirect and direct detection rates of DM and validate our predictions against the current constraints. In the neutrino sector, we run multi-dimensional scan to find data points where they can reproduce six observables from neutrino data by using the chi-square method. Also, we test these points against charged-lepton flavour violation constraints. Moreover, it is crucial to take into account the constraint from muon $g-2$ which partly excludes our parameter space. With these features, the model is unique and minimal in a way that it can provide the DM candidate, explain the neutrino data with smallest numbers of parameters and fields, satisfy the muon $g-2$ limit.

%\textcolor{blue}{With these features, the model is unique and minimal in a way that it can provide the DM candidate, explain the neutrino data with smallest numbers of parameters and fields, satisfy the muon $g-2$ limit. More importantly, the spontaneous symmetry breaking of new scalar gives mass to the DM, heavy neutrinos and relate the new vacuum expectation value (VEV) to the PQ breaking. In addition, the axion-SM sector is derived straightforwardly from the kinetic term of SM fermions, not from the effective operators.}

The paper is organised as follows. The model and particle contents are discussed in detail in section~\ref{sec:The model}. In section~\ref{sec:pheno}, we discuss the model phenomenology and relevant constraints. Numerical results are presented in section~\ref{sec:The results}. Finally, we summarise our findings and express an outlook in section~\ref{sec:Conclusion}. 
	
%%%%%%%%%%%%%%%%%%%%%%%%%%%%%%%%%%%%%%%%%
\section{The model}
\label{sec:The model}

In this work, we propose an extension to the SM with an electroweak (EW) singlet complex scalar ($\Phi$), two EW singlet {\it right-handed} heavy neutrinos ($N^i_{R}$), two EW singlet {\it left-handed heavy} neutrinos ($S^I_{L}$), a pair of fermionic dark matter ($\chi_L$ and $\chi_R$).~\footnote{Scenarios with one $N_R$ and one $S_L$, or one $N_R$ and two $S_L$ lead to two vanishing light neutrino masses. Scenario with two $N_R$ and one $S_L$ is compatible with neutrino oscillation data. However, it would require more free parameters than the scenario presented in this work.} The dark matter interaction with the SM particles is introduced through the global $U(1)_{\rm PQ}$-preserving terms where the Nambu-Goldstone boson corresponding to the broken $U(1)_{\rm PQ}$ is identified as the Axion-Like Particle (ALP). In addition, the accidental $\mathbb{Z}_2$ symmetry stabilises the DM sector in which the dark matter is odd and the other particles are even under this group. The global $U(1)_{\rm PQ}$ is assumed to be broken at a scale higher than the EW symmetry breaking scale. The relevant particle contents and their charge assignment under $SU(2)_{ \rm L}\times U(1)_{\rm Y} \times U(1)_{\rm PQ} \times \mathbb{Z}_2$ are summarised in Tab.~\ref{tab:particlesQN2}.\footnote{The remnant symmetry after the spontaneous breaking of $U(1)_{\rm PQ}$ cannot be used to stabilise the dark matter candidate. Therefore, a new symmetry group is needed.}

{
\setlength{\tabcolsep}{3pt}
\setlength{\arraycolsep}{0pt}
\begin{table}[h]
\centering
\begin{tabular}{|c|c|c|c|}
\toprule
& $SU(2)_{\rm L}$ & $U(1)_{\rm Y}$ & $U(1)_{\rm PQ}$\\
\hline
$L_L$ & $\mathbf{2}$ & $-1/2$  & $0$ \\
\hline
$e_R$ & 1  & $-1$ &  $-X_{H}$\\
\hline
$Q_L$ & $\mathbf{2}$ & $1/6$  & $0$ \\
\hline
$u_R$ & 1  & $2/3$ &  $X_{H}$\\
\hline
$d_R$ & 1  & $-1/3$ &  $-X_{H}$\\
\hline
$H$ & $\mathbf{2}$ & $+1/2$  & $X_H$\\
\hline
$N_{R}$ & $1$ & $0$ & $X_{H}$\\
\hline
$S_{L}$ & $1$ & $0$  & $X_{\Phi} + X_{H}$\\
\hline
$\chi_{L,R}$ & $1$ & $0$  & $0, -X_{\Phi}$\\
\hline
$\Phi$ & $1$ & $0$ & $X_\Phi$ \\
\bottomrule
\end{tabular}
\caption{The charge assignment under the electroweak and $U(1)_{\rm PQ}$ groups.}
\label{tab:particlesQN2}
\end{table}

Although the global symmetries such as $U(1)_{\rm PQ}$ are generally expected to be broken by quantum gravitational effects and could potentially induce decays that destabilise DM on cosmological timescales. To ensure the longevity of DM, we assume that the $\mathbb{Z}_2$ is an accidental symmetry arising from an underlying gauge symmetry at the UV scale. This ensures that the lowest-order operators mediating DM decay are sufficiently high-dimensional to yield a lifetime exceeding the cosmological age.

%--------------------------------------
\subsection{The scalar sector}
\label{sec: The scalar mixing}
In this section, we discuss the scalar sector with some detail.

The scalar potential satisfying gauge and PQ symmetries is given by
\begin{align}
    V(H,\Phi) = -\mu_H^2 H^\dagger H + \lambda_H (H^\dagger H)^2 - \mu_\Phi^2 \Phi^\dagger \Phi + \lambda_\Phi(\Phi^\dagger \Phi)^2 + \lambda_{H\Phi} (H^\dagger H)(\Phi^\dagger \Phi).
    \label{eq: the original scalar potential}
\end{align}
Here the fields $H$ and $\Phi$ are expanded as
\begin{align}
    H = \frac{1}{\sqrt{2}}\begin{pmatrix}
        \sqrt{2}G^{+} \\
        v_h + h^\prime + i G^0
    \end{pmatrix}, \qquad
    \Phi = \frac{1}{\sqrt{2}}(v_\varphi + \varphi^\prime) e^{ia /v_\varphi},
    \label{eq: the component form of scalars}
\end{align}
where $G^{\pm}$ and $G^0$ are the would-be Goldstone bosons. The $v_h$ and $v_\varphi$ are the vacuum expectation values (vevs) of $H$ and $\Phi$, respectively. Since $U(1)_{\rm PQ}$ is assumed to be broken at scale higher than the EW scale, $v_\varphi,f_a>v_h$. Note that we employ the non-linear parametrization for $\Phi$ which makes the ALP, $a$, appears explicitly as the phase. 

The fields $h'$ and $\varphi'$ mix to form the mass eigenstates through
\begin{equation}
    \begin{pmatrix}h\\ \varphi\end{pmatrix}=
    \begin{pmatrix}
        \cos{\theta}\;\;\; & \sin{\theta} \\
        -\sin{\theta}\;\;\; & \cos{\theta}
    \end{pmatrix}
    \begin{pmatrix}h'\\ \varphi'\end{pmatrix},
\end{equation}
where
\begin{equation}
    \tan(2\theta) = \frac{\lambda_{H\Phi}v_h v_\varphi}{\lambda_{\Phi}v_\varphi^2 - \lambda_{H}v_h^2}.
    \label{eq:mixingangle}
\end{equation}
Here we identify the field $h$ with the 125 GeV Higgs boson.
The masses of $h$ and $\varphi$ are given by
\begin{align}
    m_{h,\varphi}^2 = \lambda_Hv_h^2 +\lambda_{\Phi}v_\varphi^2 \mp \sqrt{(\lambda_\Phi v_\varphi^2 - \lambda_H v_h^2)^2+\lambda_{H\Phi}^2v_h^2 v_\varphi^2}\,.
\end{align}

It is more convenient to express the quartic couplings $\lambda_H$, $\lambda_\Phi$, and $\lambda_{H\Phi}$ in terms of the vevs, masses and a mixing angle. They are given by
\begin{align}
    \lambda_H &= \frac{m_h^2 \cos^2{\theta} + m_\varphi^2 \sin^2{\theta}}{v_h^2},\nonumber\\
    \lambda_\Phi &= \frac{m_h^2 \sin^2{\theta} + m_\varphi^2 \cos^2{\theta}}{v_\varphi^2},\nonumber \\
    \lambda_{H\Phi} &= \frac{(m_\varphi^2 - m_h^2)\sin{\theta}\cos{\theta}}{v_h v_\varphi}.
\end{align}
Thus, the scalar potential can be described by three free parameters: $v_\varphi$, $m_\varphi$ and $\sin{\theta}$. We note that $\sin\theta$ is constrained by the 125 GeV Higgs boson data. The combined Higgs signal strength ($\mu$) measurements from the ATLAS and CMS collaborations yields $\mu=1.03\pm0.04$~\cite{ATLAS:2016neq,CMS:2022dwd,ATLAS:2022vkf}. This results in an upper bound on the mixing angle $\sin\theta\le0.15$ at 95\% confidence level (CL). 

The minimum of the scalar potential can be found by substituting Eq.(\ref{eq: the component form of scalars}) in and setting fields to zero
\begin{eqnarray}
    V(H,\Phi)_{\rm min} = -\frac{\mu_H^2}{2}v_h^2 - \frac{\mu_\Phi^2}{2}v_\varphi^2 + \frac{\lambda_H}{4} v_h^4 + \frac{\lambda_\Phi}{4} v_\varphi^4 + \frac{1}{4} \lambda_{H\Phi} v_h^2 v_\varphi^2.
 \end{eqnarray}
This minimal potential leads to two minimization conditions:
\begin{eqnarray}
    v_h(\lambda_H v_h^2 - \mu_H^2 + \frac{1}{2} \lambda_{H\Phi}v_\varphi^2 ) = 0, \quad \text{and} \quad v_\varphi(\lambda_{\Phi}v_\varphi^2 - \mu_\Phi^2 + \frac{1}{2} \lambda_{H\Phi} v_h^2) = 0
\end{eqnarray}
In general, there are nine sets of $(v_h,v_\varphi)$ that give the minimum value of scalar potential. The positive and non-trivial one corresponds to
\begin{eqnarray}
    v_h = \sqrt{\frac{4 \mu_H^2 \lambda_\Phi - 2\mu_\Phi^2 \lambda_{H\Phi}}{4\lambda_H \lambda_\Phi - \lambda_{H\Phi}^2}}, \quad 
    v_\varphi = \sqrt{\frac{4\mu_\Phi^2 \lambda_H - 2 \mu_H^2 \lambda_{H\Phi}}{4 \lambda_H \lambda_\Phi - \lambda_{H\Phi}^2}} 
    .
\end{eqnarray}
The local minima can be verified by identifying the Hessian matrix:
\begin{eqnarray}
    \mathcal{H} = \begin{pmatrix}
        -\mu_H^2 + 3 \lambda_H v_h^2 + \frac{\lambda_{H\Phi} v_\varphi^2}{2} & \lambda_{H\Phi} v_h v_\varphi \\
        \lambda_{H\Phi} v_h v_\varphi & -\mu_\Phi^2 + 3 \lambda_{\Phi} v_\varphi^2 + \frac{\lambda_{H\Phi} v_h^2}{2}
    \end{pmatrix}.
\end{eqnarray}
By solving for the conditions that lead to positive eigenvalues in Hessian matrix~\cite{Belyaev:2022shr}, the parameters in scalar potential obey
\begin{eqnarray}
\mu_H\neq0\; \text{and}\; \mu_\Phi\neq 0\; \text{and}\;
\left\{
\begin{array}{l}
     \lambda_{H\Phi}<0\; \text{and}\; \lambda_H>0 \; \text{and}\; \lambda_\Phi>0 \; \text{and}\; \lambda_{H\Phi}^2 < 4\lambda_H \lambda_\Phi, \\
     \text{or}\\
     \lambda_{H\Phi}>0\; \text{and}\; 2\lambda \mu_\Phi^2> \lambda_{H\Phi}\mu^2\; \text{and}\; 2\lambda_\Phi \mu^2 > \lambda_{H\Phi} \mu_\Phi^2.
\end{array}
\right.
\end{eqnarray}

The ALP potential, on the other hand, is generated from the non-perturbative effect, which breaks $U(1)_{\rm PQ}$ into a shift symmetry, $a \to a + f_a$.
This shift symmetry dictates that the ALP potential takes the form of the instanton cosine potential~\cite{Adams:2022pbo}
\begin{align}
    V(a) = \widetilde{\Lambda}_a^4 \Big[ 1 - \cos{\left(\frac{2\pi a}{f_a}\right)} \Big],
    \label{eq: the ALP potential}
\end{align}
where $\widetilde{\Lambda}_a^4$ is the maximum value of the potential. By identifying this shift symmetry with $U(1)_{\rm PQ}$ transformation, one obtains a relation 
\begin{equation}
    f_a = X_{\Phi}v_\varphi \label{eq:vev and decay constant of ALP}.
\end{equation}
The ALP potential also generates the ALP mass
\begin{equation}
    m_a^2 = \frac{\Lambda_a^4}{f_a^2},
\end{equation}

where $\Lambda_a^2 \equiv 2\pi\widetilde{\Lambda}_a^2$. Since there is no direct constraint on the ALP potential, we will choose $m_a$ and $f_a$ as a free parameter of the model. 
%--------------------------------------
\subsection{The neutrino sector}
\label{sec: The neutrino masses and mixing}
In this section, we derive the light neutrino mass matrix in the light of inverse seesaw mechanism. It is well-known that the canonical seesaw mechanism predicts a high energy scale which is barely possible to test the model with existing experiments. The inverse seesaw mechanism requires more ingredient in such a way that more heavy neutrino states are needed, especially, the Dirac mass of the left- and right-heavy states which is generated from the Yukawa terms.

The gauge and PQ invariant Yukawa Lagrangian is given by
\begin{align}
    \mathcal{L}_{\rm Yuk} \supset  -Y^{\alpha i} \overline{L}_L^\alpha \tilde{H}N_{R}^i - Y^{\prime Ii} \Phi \overline{S_{L}}^I N_{R}^i  + \mathrm{h.c.} ,
    \label{eq:yukawa}
\end{align}
where $\alpha$, $i$, and $I$ are generation indices, and $\tilde{H} = i\sigma_2 H^*$. These interactions by themselves do not violate lepton number, and hence cannot generate neutrino mass. Thus, we also introduce the LNV Majorana mass terms
\begin{equation}
    \mathcal{L}_{\rm Maj.} = - \frac{\mu^{ij}_R}{2}\overline{N_{R}^{c}}^i N_{R}^j - \frac{\mu^{IJ}_S}{2} \overline{S_{L}}^I {S_{L}^{c}}^J + \mathrm{h.c.},
    \label{eq:majorana}
\end{equation}
where $\psi^c = C\overline{\psi}^T$ is the charge conjugated field. The $\mu_R$ and $\mu_S$ terms also break $U(1)_{\rm PQ}$ softly. However, they are expected to be small thanks to 't Hooft technical naturalness argument. After $U(1)_{\rm PQ}$ and EW symmetry breaking, Eqs.~(\ref{eq:yukawa}) and (\ref{eq:majorana}) lead to the neutrino mass terms \
\begin{align}
-\mathcal{L}_{\rm mass}^{\nu} =& 
    m^{\alpha i}_D  \overline{\nu^\alpha_{L}}
    N_R^i + M^{Ij}_N \overline{S_{L}}^I N_R^j  \nonumber\\
    & + \frac{\mu_R^{ij}}{2}\overline{N_{R}^c}^i {N_{R}}^j + \frac{\mu_S^{IJ}}{2}\overline{S_{L}}^I {S_{L}^c}^J + \mathrm{h.c.} \label{eq: neutrino mass term},
\end{align}
where we define $m_D^{\alpha i} = Y^{\alpha i }v_h/\sqrt{2}$ and $M_{N}^{Ij} = Y^{\prime Ij} v_{\varphi}/\sqrt{2}$.
The neutrino mass terms can be written in a compact form
\begin{align}
-\mathcal{L}_{\rm mass}^{\nu} 
    =& \frac{1}{2}\overline{\mathcal{N}^c_R}\mathcal{M}_\nu\mathcal{N}_R + \mathrm{h.c.},
\end{align}
where 
\begin{align}
\mathcal{N}_R = \begin{pmatrix}
    \nu_L^c \\
    N_R \\
    S_{L}^c
\end{pmatrix},
\qquad
    \mathcal{M}_\nu = \begin{pmatrix}
        \mathbf{0} & m_D & \mathbf{0} \\
        m^T_D & \mu_R & M^T_N \\
       \mathbf{0} & M_N & \mu_S
    \end{pmatrix}.
    \label{eq: the mass matrix of the model}
\end{align}

The neutrino mass matrix $\mathcal{M}_\nu$ can be brought to a block diagonal form by~\cite{Dev:2012sg}
\begin{align}
    V^T 
    \mathcal{M}_\nu V 
    = \begin{pmatrix}
        (M_{\rm light})_{3\times3} & \bf0\\
        \bf 0 & (M_{\rm heavy})_{4\times4}
    \end{pmatrix},
\end{align}
where $V$ is a unitary matrix, $M_{\rm light}$ and $M_{\rm heavy}$ are the light neutrino and heavy neutrino mass matrices, respectively. In the limit where $||\mu_R||,||\mu_S||\ll||m_D||\ll||M_N||$~\footnote{Here we define $||M|| \equiv \sqrt{{\rm Tr}(M^\dagger M)}$ .}, $M_{\rm light}$ can be approximated, in $(||m_D||/||M_{N}||)^2$, by~\cite{Pilaftsis:2008qt}
\begin{equation}
    %M_{\rm light} \simeq m_D \left(\mu_R - M_N^T\mu_S^{-1} M_N\right)^{-1}m_D^T. \label{eq: light neutrino mass}
    M_{\rm light} \simeq -m_D M_N^{-1}\mu_S (M_N^T)^{-1}m_D^T. 
    \label{eq: light neutrino mass}
\end{equation}
Also, $M_{\rm heavy}$ is given by~\cite{Pilaftsis:2008qt}
\begin{eqnarray}
    M_{\rm heavy} \simeq 
    \begin{pmatrix}
        \mu_R & M_N^T \\
        M_N & \mu_S
    \end{pmatrix} .
    \label{eq: heavy neutrino mass}
\end{eqnarray}
The rotation matrices for $M_{\rm light}$ and $M_{\rm heavy}$ is defined as 
\begin{eqnarray}
    U_P^T M_{\rm light} U_P = {\rm diag}.(m_1,m_2,m_3), \quad U_R^T M_{\rm heavy} U_R = {\rm diag}.(M_1,M_2,M_3, M_4).
\end{eqnarray}
The detailed discussion of neutrino mixing is given in appendix~\ref{sec: full neutrino mixing matrix}. The mixing matrices will play an important role in determining the neutrino oscillation observables and the cLFV appendix~\ref{sec: the mu to e gamma rate}. Note that, to lowest order, the Majonana mass term $\mu_R$ does not contribute to $M_{\rm light}$. So, we will set it to be zero in this work. Notice also that in this case, $M_{\rm light}$ receives an extra suppression of order $||\mu_S||/||M_N||$ compared to the usual seesaw scenario.  This makes it possible to place a seesaw scale within the reach of collider experiment.

%--------------------------------------
\subsection{Dark matter sector}
\label{sec: The dark matter sector}
The dynamics of the DM sector is governed by the Yukawa interaction 
\begin{align}
    \mathcal{L}_{\rm DM} = - g_{\chi}\Phi \overline{\chi_L}\chi_R + \mathrm{h.c.} .
    \label{eq: axion and dark matter interaction}
\end{align}
The $U(1)_{\rm PQ}$ symmetry breaking generates a DM mass, $m_\chi = g_\chi v_\varphi/\sqrt{2}$. Below the $U(1)_{\rm PQ}$ breaking scale, DM interactions are given by
\begin{align}
    \mathcal{L}_{\rm DM} \supset& -\frac{m_\chi}{v_\varphi}\sin{\theta}\;h \overline{\chi}\chi - \frac{m_\chi}{v_\varphi}\cos{\theta}\;\varphi \overline{\chi}\chi - i  \frac{m_\chi}{v_\varphi}\; a\overline{\chi}\gamma_5 \chi,
    \label{eq:ALPmediation}
\end{align}
where $\theta$ is the neutral scalars mixing angle defined in Eq.~\eqref{eq:mixingangle}.

From the above interaction, one sees that the $h$, $\varphi$ and $a$ can serve as a portal between the DM sector and the SM sector. However, the $h$ and $\varphi$ portals to SM fermions and gauge bosons are suppressed, compared to the a mediating diagrams, by $\sigma_{\chi \bar{\chi} \to h,\varphi \to f \bar{f}}/\sigma_{\chi \bar{\chi} \to a \to f \bar{f}} \sim \sin^2{\theta}\cos^2{\theta} = 0.014$ due to the mixing in the neutral scalar sector. The ALP portal, on the other hand, does not suffer from this suppression. Thus, DM phenomenology is solely mediated by the ALP portal. 

\subsection{The ALP-SM particle interactions}
\label{sec:the axion-fermion interaction}

% With this approach, the effective $C_{ff}$ coupling is determined by the PQ charge of fermion which is equal to that of the Higgs doublet (see, Tab.~\ref{tab:particlesQN2}). } %\textcolor{red}{CP: To Nakorn, we probably need to move this paragraph into the model section.}
In our model, the ALP interaction with SM sector is not generated from the effective operators. Instead, the kinetic terms of SM fermion, charged under PQ symmetry, can lead to the ALP-fermion coupling, while the ALP-photon interaction is also derived from the triangle loop of fermion.

The ALP-fermion interaction can be derived by demanding that the Lagrangian of the model is invariant under the following PQ transformation
\begin{eqnarray}
f_L &\mapsto& f_L, \quad   f_R \mapsto e^{-iX_{f}\; a/f_a} f_R, \quad  \psi \mapsto e^{-iX_{\psi}\; a/f_a} \psi,\quad
    \eta \mapsto e^{-iX_{\eta} a/f_a}\eta,
    \label{eq:PQ transformation on general fermions}
\end{eqnarray}
where $f$ is the SM fermion, $\psi = \{ \chi,  N\}$ and $\eta=\{H,\Phi\}$. The $X_f,X_\psi$ and $X_{\eta}$ are PQ-charges of SM fermions, additional fermions and scalars respectively. According to Tab.~\ref{tab:particlesQN2}, the PQ transformation, Eq.~(\ref{eq:PQ transformation on general fermions}), on the kinetic term of SM fermions leads to ALP-SM fermion derivative interactions in the form of
\begin{eqnarray}
    \mathcal{L}_{\rm kin} &\supset& \overline{f} i\gamma^\mu \partial_\mu f \overset{\rm PQ\;trans.}{\longrightarrow}\overline{f} i\gamma^\mu \partial_\mu f + C_{ff}\frac{\partial_\mu a}{f_a} \overline{f}\gamma^\mu \gamma_5 f,
    \label{eq: L kin of fermion}
\end{eqnarray}
where $C_{ff}= X_{f} = \pm X_{\eta}$ where $+$ is for up-type quarks and $-$ is for charged lepton and down-type quarks, see a possible set of PQ charge in Tab.~\ref{tab:particlesQN2}.

\begin{figure}[t]
\centering
\begin{subfigure}[b]{0.4\textwidth}
\includegraphics[width=\textwidth]
{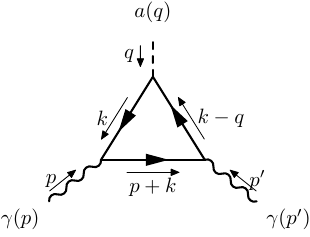}
\centering\small{(a)}
\end{subfigure}
\begin{subfigure}[b]{0.4\textwidth}
\includegraphics[width=\textwidth]
{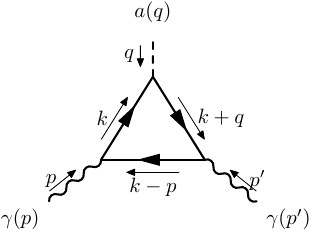}
\centering\small{(b)}
\end{subfigure}
\caption{The loop diagrams for $a\gamma\gamma$. The particles circulating in the triangle loops are quarks ($t,b,c,s,u,d$) and charged leptons ($e, \mu, \tau$).}
\label{fig: triangular loop}
\end{figure}

%While the axion-photon interaction is also derived from the triangle loop of fermion.  

The coupling of ALP-photon is generated by the triangle diagrams as shown in Fig.~\ref{fig: triangular loop} where the vertex function is given by
\begin{eqnarray}
    i \Pi_{(a)}^{\mu\nu} =&& (-1) \int \frac{d^4k}{(2\pi)^4} {\rm Tr}\Bigg[\left(-ie Q_f \gamma^\mu\right) \left[\frac{i(\slashed{k}-\slashed{p}+m_f)}{(k-p)^2 - m_f^2} \right] \left(-ie Q_f \gamma^\nu\right) \left[ \frac{i(\slashed{k}+\slashed{q}+m_f)}{(k+q)^2-m_f^2} \right] \nonumber\\
    &&\left(-i \frac{C_{ff}}{f_a} \slashed{q} \gamma_5 \right) \left[\frac{i(\slashed{k}+m_f)}{k^2-m_f^2}  \right]
    \Bigg],\\
    i \Pi_{(b)}^{\mu\nu} =&& (-1) \int \frac{d^4 k}{(2\pi)^4} {\rm Tr}\Bigg[
    \left(-ie Q_f \gamma^\mu\right) \left[\frac{i(\slashed{k}+m_f)}{k^2 - m_f^2} \right] \left( -i\frac{C_{ff}}{f_a} \slashed{q}\gamma_5 \right) \left[ \frac{i(\slashed{k}-\slashed{q}+m_f)}{(k-q)^2 - m_f^2} \right] \nonumber\\
    && \left(-ieQ_f \gamma^\nu\right) \left[ \frac{i(\slashed{k}+\slashed{p}+m_f)}{(k+p)^2 - m_f^2}  \right] 
    \Bigg],
\end{eqnarray}
where the total vertex function is the sum of two functions: $i\Pi^{\mu\nu} = i\Pi^{\mu\nu}_{(a)}+i\Pi^{\mu\nu}_{(b)}$. In the limit that the mass of fermion in the loop is much less than the Center-of-Mass (CM) energy, $\sqrt{s}$, we find that
\begin{eqnarray}
    i\Pi^{\mu\nu} =&& \sum_{f} 2 \log{(2)^2}\left( \frac{m_f^2}{s}\right)\left(\frac{C_{ff}}{f_a} \right) \alpha_{\rm EM}  \epsilon^{\mu\nu \rho \sigma}  p_{\rho} p^\prime_{\sigma},
\end{eqnarray}
where $f$ stands for quarks $(t,b,c,s,u,d)$ and charged leptons $(e,\mu,\tau)$. The $p$ and $p^\prime$ are the momenta of incoming photons and $\alpha_{\rm EM}$ is the fine structure constant.

%%%%%%%%%%%%%%%%%%%%%%%%%%%%%%%%%%%%%%%%%
\section{Phenomenology}
\label{sec:pheno}
In this section, we study the phenomenology of our model as discussed in section~\ref{sec:The model}. This contains the discussion on 1) neutrino masses and mixing, 2) dark matter, and 3) the muon $g-2$ measurement consistently.

%-------------------------------
\subsection{Neutrino masses and mixing}
\label{sec:numassandmixing}
In our scenario $m_D$ is a complex $3\times2$ matrix, $M_N$ is a complex $2\times2$ matrix and $\mu_S$ is a symmetric complex $2\times2$ matrix. However, not all components of these matrices are physical. One can perform unitary field redefinitions on $N_R^i$ and $S_L^I$ to put $M_N$ and $\mu_S$ in the form
\begin{eqnarray}
    M_N &=& \bar M_N \begin{pmatrix}
        1\;\;\;\; & 0 \\
        0\;\;\;\; & 1
    \end{pmatrix},
    \label{eq:mN}\\
    \mu_S &=& \bar \mu_S \begin{pmatrix}
        1\;\;\;\; &\alpha\\
        \alpha\;\;\;\; &\beta\, e^{i\phi}
    \end{pmatrix},
\end{eqnarray}
where $\alpha$, $\beta$ and $\phi$ are real and positive. The $\bar M_N = Y^\prime v_\varphi/\sqrt{2}$ and $\bar \mu_S$ are the overall scale of matrix $M_N$ and $\mu_S$, respectively.\footnote{The equivalent parameterisation of $M_N$ and $\mu_S$ are given by
\begin{eqnarray}
    M_N = \bar M_N \begin{pmatrix}
        1\;\;\; & 0 \\
        0\;\;\; & \gamma
    \end{pmatrix}, \quad 
    \mu_S = \bar \mu_S \begin{pmatrix}
        1\;\;\; & \alpha \gamma \\
        \alpha \gamma\;\;\; & \beta \gamma^2 e^{i \phi}
    \end{pmatrix}.
\end{eqnarray}
However, $\gamma$ does not play a role in the analysis. Therefore, we will remove it from $M_N$ and $\mu_S$ matrices.} Moreover, by redefining the phases of $L^\alpha$ and $N^i$, one can rotate away 4 phases in $m_D$. From our analysis, we find two minimal textures of $m_D$, which lead to viable neutrino masses and mixing. They are
\begin{align}
    m_D^{(A)} &= \bar m_D \begin{pmatrix}
        1\;\;\;\; & 0\\
        y_1 \;\;\;\;&y_2\\
        0\;\;\;\;&y_3
    \end{pmatrix},\label{eq:mDA}\\
    m_D^{(B)} &= \bar m_D \begin{pmatrix}
        1\;\;\;\; &0 \\
        0\;\;\;\;&y_1\\
        y_2\;\;\;\;&y_3
    \end{pmatrix}\label{eq:mDB},
\end{align}
where $\bar m_D = Y v_h/\sqrt{2}$. Note that the Yukawa coupling $Y,y_1Y,y_2Y, y_3Y\leq \sqrt{4\pi}$. Note that $m_D$ with only two non-vanishing elements leads to 2 massless neutrinos and hence is incompatible with neutrino oscillation data. Among the remaining structures, a systematic scan over all textures with three free parameters $(y_1,y_2,y_3)$ shows that only Eqs. (\ref{eq:mDA}) and (\ref{eq:mDB}) reproduce the six oscillation observables within the $3\sigma$ ranges of Tab. \ref{tab:the neutrino oscillation observables}.

Finally, the minimal mass matrix (Eq.(\ref{eq: light neutrino mass})) for light neutrinos takes one of the following form:
\begin{align}
    M_{\rm light}^{(A)} &= \kappa \begin{pmatrix}
        1\;\;\; && y_1+\alpha\,y_2 \;\;\; &&\alpha\,y_3\\
        y_1+\alpha\,y_2 \;\;\; &&y_1^2+2\alpha\,y_1y_2+\beta e^{i\phi}y_2^2 \;\;\; &&\alpha\,y_1y_3 + \beta e^{i\phi}y_2y_3\\
        \alpha\,y_3 \;\;\; &&\alpha\,y_1y_3 + \beta e^{i\phi}y_2y_3 \;\;\; &&\beta e^{i\phi}y_3^2
    \end{pmatrix}\label{eq:MA light},\\
    M_{\rm light}^{(B)} &= \kappa \begin{pmatrix}
        1\;\;\;&& \alpha\,y_1 \;\;\; &&y_2+\alpha\,y_3\\
        \alpha\,y_1 \;\;\; &&\beta e^{i\phi}y_1^2 \;\;\; &&\alpha\,y_1y_2 + \beta e^{i\phi}y_1y_3\\
        y_2+\alpha\,y_3 \;\;\;&&\alpha\,y_1y_2 + \beta e^{i\phi}y_1y_3 \;\;\;  &&y_2^2+2\alpha\,y_2y_3+\beta e^{i\phi}y_3^2
    \end{pmatrix},
    \label{eq:MB light}
\end{align}
where the prefactor $\kappa$ is defined as
\begin{eqnarray}
    \kappa = \frac{Y^2}{Y^{\prime 2}}\frac{v_h^2}{v_\varphi^2}\bar \mu_S.
    \label{eq: kappa}
\end{eqnarray}
Thus, in our scenario, $M_{\rm light}$ depends on seven real parameters: $\kappa$, $y_{1,2,3}$, $\alpha$, $\beta$, and $\phi$.

{
\setlength{\tabcolsep}{5pt}
\setlength{\arraycolsep}{5pt}
\begin{table}[ht!]
\centering
\begin{tabular}{|c||c|c|}
\toprule
{\bf Parameters} & {\bf Normal Ordering} & {\bf Inverted Ordering} \\
\midrule
$\sin^2{\theta_{12}}$ & $0.308^{+0.037}_{-0.033}$ & $0.308^{+0.037}_{-0.033}$\\
$\sin^2{\theta_{13}}$ & $0.02215^{+0.00173}_{-0.00185}$ & $0.02231^{+0.00178}_{-0.00171}$\\
$\sin^2{\theta_{23}}$ & $0.470^{+0.115}_{-0.035}$ & $0.550^{+0.034}_{-0.110}$\\
$m^2_{\rm solar}/10^{-5}$ eV$^{-2}$ & $7.49^{+0.56}_{-0.57}$ & $7.49^{+0.56}_{-0.57}$\\
$m_{\rm atm}^2/10^{-3}$ eV$^{-2}$ & $2.513^{+0.065}_{-0.062}$ & $2.484^{+0.063}_{-0.063}$\\
$\delta_{\rm CP}/\degree$ & $212^{+152}_{-\phantom{1}88}$ & $274^{+61}_{-73}$\\
\bottomrule
\end{tabular}
\caption{\label{tab:the neutrino oscillation observables}The best fit values of neutrino oscillation parameters and their $3\sigma$ ranges as determined from the global neutrino data~\cite{Esteban:2024eli}. Here $m^2_{\rm solar}= m_2^2-m_1^2$ and $m_{\rm atm}^2 = m_3^2-m_1^2$ ($m_2^2-m_3^2$) for normal ordering (inverted ordering).}
\label{tab:neutriodata}
\end{table}
}

The light neutrino mass matrix can be diagonalized by a unitary matrix $U_{\rm PMNS}$,
\begin{equation}
    U^T_{\rm PMNS} M_{\rm light}U_{\rm PMNS} = \text{diag.}(m_{1},m_{2},m_{3}),
\end{equation}
where $m_{i}>0$ are the mass eigenvalues. The above equation defines the Pontecorvo–Maki –Nakagawa–Sakata (PMNS) matrix. In general, the PMNS matrix contains three mixing angles, one Dirac phase, and two Majorana phases. 

Once the neutrino mass matrix $M_{\rm light}$ has been diagonalized, the three mixing angles can be determined by
\begin{equation}
    s_{13}^2 = |U_{e3}|^2,\quad
    s_{12}^2 = \frac{|U_{e2}|^2}{1-|U_{e3}|^2},\quad
    s_{23}^2 = \frac{|U_{\mu3}|^2}{1-|U_{e3}|^2},
\end{equation}
where $s_{ij} = \sin\theta_{ij}$ and $U\equiv U_{\rm PMNS}$.
The Dirac phase, $\delta_{\rm CP}$, is encoded in the Jarsklog invariant
\begin{equation}
    J = \text{Im}\left(U_{\mu3}^{\phantom{*}}U_{e3}^*U_{e2}^{\phantom{*}}U_{\mu2}^*\right)= c_{12} s_{12} c_{23} s_{23} c_{13}^2 s_{13} \sin{\delta_{\rm CP}} .
\end{equation}
The remaining two Majorana phases, in principle, can be deduced from the invariants Im[$(U_{e2}^{\phantom{*}}U_{e1}^*)^2$] and Im[$(U_{e3}^{\phantom{*}}U_{e1}^*)^2$]~\cite{Jenkins:2007ip}. However, both Majorana phases have never been determined experimentally. The values of neutrino masses, mixing angles, and the Dirac phase, as determined from the global analysis of neutrino data, are collected in Tab.~\ref{tab:neutriodata}. 

An interesting feature of our set up is that the light neutrino mass matrix is rank 2. As a result, the lightest neutrino mass is vanishing. Hence, the other two non vanishing neutrino masses depend only on $\Delta m^2_{\rm sol}$ and $\Delta m^2_{\rm atm}$. In particular, the sum of neutrino masses is given by $\sum_im_i = \sqrt{\Delta m^2_{\rm sol} } + \sqrt{\Delta m^2_{\rm atm} }$ and $\sqrt{\Delta m^2_{\rm atm} - \Delta m^2_{\rm sol} } + \sqrt{\Delta m^2_{\rm atm} }$ for the NO and IO scenarios respectively \footnote{Note that we are considering a minimal scenario where they are only 2 RHNs. For more complicated models, the above constraint does not apply.}. The sum of neutrino masses is constrained by cosmological measurement. Combining the Planck 2018 cosmic microwave background data and the baryonic acoustic oscillation measurement result in an upper bound $\sum_im_i\le120$ meV at 95\% CL~\cite{Planck:2018vyg}. In addition to the neutrino masses sum, there are two effective masses relevant for laboratory based neutrino experiments. First, the effective mass $m_{\nu_e}^{\rm eff} = \sqrt{\sum_i|U_{ei}|^2m_i^2}$ determines the endpoint of beta decay spectrum. The latest measurement published by KATRIN has put an upper bound of $m_{\nu_e}^{\rm eff}\le 450$ meV at 90\% CL~\cite{Katrin:2024tvg}. Second the rate for neutrinoless double beta ($0\nu\beta\beta$) decay depends on the effective mass $m_{ee} = |\sum_i U_{ei}^2m_i|$. The strongest upper bound on $m_{ee}$ is provided by the KamLAND-Zen experiment with $m_{ee}\le36-156$ meV at 90\% CL~\cite{KamLAND-Zen:2022tow}.

%-------------------------------
\subsection{Dark matter phenomonology}
\label{sec:The dark matter constraints}

In addition to generating neutrino masses, our model also provides a DM candidate which is assumed to be a fermion. In the absence of the neutral scalar mixing, $\theta$, DM interacts with the SM fermions purely through the ALP, see Eqs.~(\ref{eq:ALPmediation}) and~(\ref{eq: L kin of fermion}). Moreover, this model gives additional contributions to the muon $g-2$ which is constrained by BNL+FNAL measurements from Run 1-6, see section~\ref{sec:The anomalous magnetc moment of muon}. 

\subsection*{The DM relic density}
The energy density of DM is precisely measured by the Planck satellite~\cite{Planck:2018vyg} to be $\Omega_{\chi}^{\rm Planck} h^2 = 0.1200 \pm 0.0012$. 
%\begin{align}
%    \Omega_{\chi}^{\rm Planck} h^2 = 0.12 \pm 0.0012.
%    \label{eq: planck relic density}
%\end{align}
Assuming DM is a thermal relic, we can compute its density by assuming that DM was in thermal equilibrium with the thermal bath in the early Universe and underwent thermal freeze-out. The evolution of DM density is governed by the Boltzman equation
\begin{eqnarray}
    \frac{d n_{\chi}}{dt} + 3 H n_{\chi} = -\braket{\sigma v_{\rm rel}} \Big[n_{\chi}^2 - (n_{\chi}^{\rm eq})^2\Big],
    \label{eq: BE for wimp}
\end{eqnarray}
where $H$ is the Hubble constant, $\langle\sigma v_{\text{rel}}\rangle$ is the thermal average DM annihilation cross-section, and $n_\chi(n_\chi^{(\rm eq)})$ is the (equilibrium) number density of DM. Even though the Boltzman equation can be solved numerically, in this work we will adopt the simple freeze-out approximation. According to the freeze-out approximation, the initial DM density $n_\chi$ is assumed to follow $n_\chi^{\rm eq}$ until the freeze-out temperature is reached. After that $n_\chi$ becomes frozen while $n_\chi^{\rm eq}$ drops exponentially. The temperature at which $n_\chi$ freezes out is determined by~\cite{Kolb:1990vq}
\begin{eqnarray}
    x_{f} = \log{\left( \frac{0.038 g_{\chi}m_{\rm Pl}m_{\chi}\braket{\sigma v_{\rm rel}}}{g_*^{1/2} x_f^{1/2}}\right)},
\end{eqnarray}
where $g_\chi=4$ is the DM internal degrees of freedom, $m_{\rm Pl} = 1.22\times10^{19}$ GeV is the Planck mass, and $g_{*}$ is the relativistic degree of freedoms at the time of freeze-out. 

The present DM energy density can be approximated by 
\begin{eqnarray}
    \Omega_{\rm DM}h^2 = \frac{1.07\times 10^{9}\;\; {\rm GeV}^{-1}}{g_*^{1/2}m_{\rm Pl}}\frac{x_f}{\braket{\sigma 
    v_{\rm rel}}}.
\end{eqnarray}
The above approximation reproduces the Planck observation for $\langle\sigma v_{\rm rel}\rangle\simeq2.6\times10^{-26}$ cm$^3$/s, the so-called thermal relic cross-section. It should be note that our thermal relic cross-section lies in the middle of the precise calculation $\langle\sigma v_{\rm rel}\rangle=2.2\times 10^{-26}$ cm$^3$/s of Ref.~\cite{Steigman:2012nb} and the canonical value $\langle\sigma v_{\rm rel}\rangle=3.0\times 10^{-26}$ cm$^3$/s commonly quoted in the literature. Thus, following Ref.~\cite{Treesukrat:2024akz}, we will take the range $(2.2-3.0)\times10^{-26}$ cm$^3$/s as our thermal relic cross-section to reflect the uncertainty in the computation.

\begin{figure}[thb!]
\centering
\begin{subfigure}[b]{0.4\textwidth}\includegraphics[width=\textwidth]{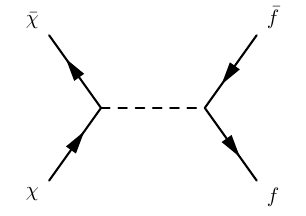}
\centering\small{(a) leading diagram}
\end{subfigure}
\begin{subfigure}[b]{0.43\textwidth}\includegraphics[width=\textwidth]{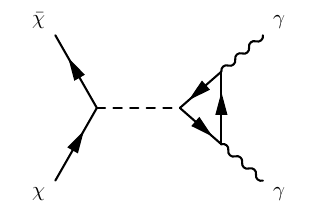}
\centering\small{(b) sub-leading diagram}
\end{subfigure}
\caption{The relevant Feynman diagrams for DM annihilation into a pair of SM fermions (a) and photons (b). Since the $\chi \bar\chi \to \gamma\gamma$ contributes less than 1\% to the relic density, we consider only the $\chi\chi \to f \bar f$}
\label{fig:the DM annihilate into ff}
\end{figure}

In our scenario we make a simplifying assumption that $\varphi$, $a$ and $\nu_{\rm heavy}$ are heavier than $\chi$ so that DM can only annihilate into SM particles. Also, as we have shown in section \ref{sec: a triangular loop function} that $\langle \sigma v \rangle_{\gamma\gamma}\propto \alpha_{\rm EM}^2$ . We can neglect the contribution from the loop to the relic density and DM indirect detection. Therefore, the relevant annihilation channels is $\bar\chi\chi \to \bar f f$, see Fig.~\ref{fig:the DM annihilate into ff}(a) where the cross section is
\begin{align}
    \langle\sigma_{f\overline{f}}v_{\rm rel}\rangle &\simeq 
    \frac{8}{\pi} \frac{C_{ff}^2 X_{\Phi}^4 }{f_a^4}\frac{m_f^2m_\chi^4}{(m_a^2 - 4m_\chi^2)^2} \sqrt{1-\frac{m_f^2}{m_\chi^2 }}
    \label{eq:dmannihilationmumu},
\end{align}
while the cross section for Fig.~\ref{fig:the DM annihilate into ff}(b) is
\begin{eqnarray}
    \langle \sigma_{\gamma\gamma} v_{\rm rel}\rangle = \frac{\log{(2)}^4}{8\pi }\frac{C_{ff}^2 X_\Phi^2}{f_a^4}\alpha_{\rm EM}^2 \frac{m_f^4 m_{\chi}^2}{(m_a^2 - 4m_\chi^2)^2}.
    \label{eq:dmannihilationgammagamma}
\end{eqnarray}
Here $m_f$ and $m_a$ are the SM fermion and ALP masses, respectively. The $C_{ff}$ is the ALP-SM fermion coupling and $X_\Phi$ is the PQ charge of $\Phi$ and equal to $f_a/v_\varphi$. Notice the $\langle \sigma_{f \bar f}v \rangle \propto m_f^2$. It makes sense to consider only the $\chi \bar \chi \to t \bar t$ process since the lighter quark contribution to the annihilation cross section is suppressed by $m_b^2/m_t^2 \sim (4.2/173)^2 \sim 6 \times 10^{-4}$. The $\chi \bar \chi \to \gamma\gamma$ cross section is also sub-leading compared to the $\chi\bar\chi \to t \bar t$ where $\langle \sigma_{\gamma\gamma} v_{\rm rel} \rangle/\langle \sigma_{t \bar t} v_{\rm rel} \rangle \sim \alpha_{\rm EM}^2 m_f^2/m_\chi^2$ up to some numerical factor. For DM mass around 1 TeV, $\langle \sigma_{\gamma\gamma} v_{\rm rel} \rangle/\langle \sigma_{t \bar t} v_{\rm rel} \rangle \sim 10^{-9}$ which is negligible when we compute the relic density and DM indirect detection. However, for completeness, we present the plot of $\langle \sigma_{\gamma\gamma} v_{\rm rel} \rangle$ and constrain it with the current experimental limit in Fig.~\ref{fig: dm dm to photon photon}.

\subsection*{The DM direct detection}
\label{sec: DM-nucleon scattering cross section}

\begin{figure}[tbh!]
\centering
\begin{subfigure}[b]{0.4\textwidth}
\includegraphics[width=\textwidth]
{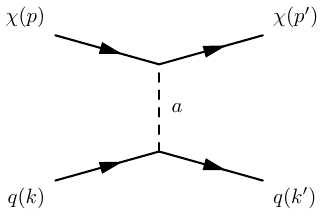}
\centering\small{(a) quark level diagram}
\end{subfigure}
\begin{subfigure}[b]{0.4\textwidth}
\includegraphics[width=\textwidth]
{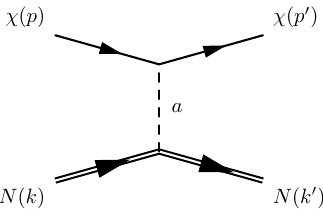}
\centering\small{(b) nucleon level diagram}
\end{subfigure}
\caption{The diagrams for quark- and nucleon-DM elastic scattering.}
\label{fig: direct detection diagrams}
\end{figure}

In this section, we compute the elastic scattering cross section of DM on nucleon and we demonstrate that the cross section is far below the exclusion limit from the Lux-Zeplin (2025) ~\cite{LZ:2024zvo} due to the velocity suppression.

We will closely follow the calculation in Ref.~\cite{Lin:2019uvt}. In our model, the effective operator of quark-DM scattering in the limit of small momentum transfer reads
\begin{eqnarray}
    \mathcal{M} = \left(-\frac{m_\chi C_{ff}}{v_\varphi f_a m_a^2}\right) \big( \bar \chi \gamma_5 \chi \big) \big(\bar q (\slashed{p}-\slashed{p}^\prime)\gamma_5 q\big).
    \label{eq:quark level amplitude}
\end{eqnarray}
The corresponding nucleon matrix element to Eq.(\ref{eq:quark level amplitude}) is given by
\begin{eqnarray}
    \mathcal{M} &&= \left(-\frac{m_\chi C_{ff}}{v_\varphi f_a m_a^2}\right) \bra{\chi (p^\prime)} \bar \chi \gamma_5 \chi \ket{\chi(p)} \bra{N(k^\prime)} \bar q (\slashed{p}-\slashed{p}^\prime)\gamma_5 q \ket{N(k)}\nonumber \\
    && \simeq \left(\frac{m_\chi C_{ff}}{v_\varphi f_a m_a^2}\right) \Delta_q^N [\bar u_{\chi}(p^\prime) \gamma_5 u_{\chi}(p)] [\bar u_N(k^\prime)(\slashed{k}-\slashed{k}^\prime)\gamma_5 u_N(k)].
\end{eqnarray}
where we have dropped the term of $q_{\mu}/m_N$.
Making use of the Dirac solution in NR limit 
\begin{eqnarray}
    u_s(k) \simeq 
    \begin{pmatrix}
        \left(\sqrt{m} - \frac{1}{2} \frac{\vec\sigma\cdot \vec k}{\sqrt{m}} + \cdots \right) \xi_s \\
        \left(\sqrt{m} + \frac{1}{2} \frac{\vec\sigma\cdot \vec k}{\sqrt{m}} + \cdots \right) \xi_s
    \end{pmatrix},
\end{eqnarray}
where $\xi_\uparrow=(1,0)^T$, $\xi_\downarrow=(0,1)^T$ are 2-spinors for spin up and down, respectively. By using the NR solution, we find, up to the leading order of the expansion,
\begin{eqnarray}
    \bar \chi \gamma_5 \chi  \simeq \xi^\dagger \vec \sigma\cdot(\vec k - \vec k^\prime) \xi,\;
    \bar q \gamma^0 \gamma_5 q \simeq \xi^\dagger \sigma\cdot(\vec k + \vec k^\prime) \xi,\;
    \bar q \gamma^i \gamma_5 q \simeq 2m_n \xi^\dagger S^i \xi \label{eq: spinor product}
\end{eqnarray}

With the help of Eq.~(\ref{eq: spinor product}) and defining $\vec S = \vec \sigma/2$, we can write
\begin{eqnarray}
    \mathcal{M} &&= \left(\frac{m_{\chi}}{v_\varphi}\right) \left(\frac{C_{ff}}{f_a}\right)\frac{4m_n \Delta^N_q}{m_a^2} \Big(\xi^\dagger_\chi \vec S \cdot (\vec k - \vec k^\prime) \xi_\chi \Big) \Big(  \xi^\dagger_n \vec S \cdot (\vec k - \vec k^\prime) \xi_n \Big),
\end{eqnarray}
and the squared amplitude is 
\begin{eqnarray}
    \overline{|\mathcal{M}|^2} = \left(\frac{m_\chi}{v_\varphi}\right)^2 \left(\frac{C_{ff}}{f_a}\right)^2 \left( \frac{4 m_n \Delta^N_q}{m_a^2} \right)^2 \left(\frac{1}{12}\vec q^{\,2}\right)^2
\end{eqnarray}

Let us define 
\begin{eqnarray}
    b_n^2 = \frac{\overline{|\mathcal{M}|^2}}{(4m_\chi m_n)^2} = \left( \frac{C_{ff}}{v_\varphi f_a}\right)^2 \left(\frac{\Delta_q^N}{m_a^2} \right)^2 \left(\frac{1}{12} \vec q^{\,2}\right)^2
\end{eqnarray}

The scattering cross section is given by~\cite{Lin:2019uvt}
\begin{eqnarray}
    \sigma_{\chi N} = \frac{\mu_{\chi N}^2 b_n^2}{\pi} 
    = \frac{1}{144 \pi} \left(\frac{C_{ff}}{v_\varphi f_a}\right)^2 \left(\frac{1}{m_a^2}\right)^2 (\Delta_q^N)^2 \mu_{\chi N}^6 v^4.
\end{eqnarray}
Apparently, the DM-nucleon scattering cross section is suppressed by $v^2$ where $v\sim 10^{-3}$ as we can see from Fig.~\ref{fig: direct detection limit}.

\subsection*{The DM indirect detection}
The DM indirect detection is the study of the annihilation of a pair of DMs into a pair of SM particles. The processes can be depicted in Fig.~\ref{fig:the DM annihilate into ff} and the corresponding cross sections are the same as Eq.(\ref{eq:dmannihilationmumu}) and Eq.(\ref{eq:dmannihilationgammagamma}). The relevant plot of DM indirect detection is presented in Fig.\ref{fig: indirect detection limit}.

The numerical studies of parameter space and relevant constraints in DM studies will be investigated in detail in section~\ref{sec:The results}.  

%--------------------------------------
\subsection{The measurement of muon magnetic moment}
\label{sec:The anomalous magnetc moment of muon}
The recent report on the measurement of muon magnetic moment~\cite{Aliberti:2025beg,Muong-2:2025xyk} has been confirmed that the measured value agrees with the SM prediction 
\begin{align}
    \Delta a_\mu \equiv a_\mu^{\rm EXP} - a_\mu^{\rm SM} = (39 \pm 64 )\times 10^{-11},
    \label{eq:The numerical value of muon g-2}
\end{align}
where $a_\mu = (g-2)/2$. There is no tension between measurement and the SM prediction now. We can use this constraint to find the allowed parameter space on $C_{ff}/f_a$.

\begin{figure}[thb!]
\centering
\begin{subfigure}[b]{0.325\textwidth}\includegraphics[width=\textwidth]{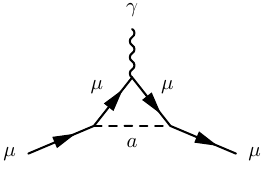}
\centering\small{(a)}
\end{subfigure}
% \begin{subfigure}[b]{0.325\textwidth}\includegraphics[width=\textwidth]{fig/diagram_g2_amugamma.pdf}
% \centering\small{(b)}
% \end{subfigure}
\begin{subfigure}[b]{0.325\textwidth}\includegraphics[width=\textwidth]{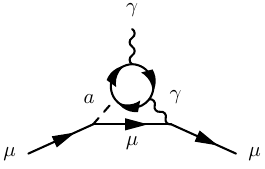}
\centering\small{(b)}
\end{subfigure}
\caption{The representative diagrams for axion contribution to muon $g-2$. The left and right panels are one-loop and two-loop contributions, respectively.}
\label{fig:The axion contribution to muon g2}
\end{figure}

In this model, the additional contributions to $a_\mu$ can be induced by ALP-SM fermion couplings 
\begin{align*}
    \mathcal{L} \supset  \frac{C_{ff}}{f_a}(\partial_\mu a) \overline{f} \gamma^\mu \gamma_5 f,
\end{align*}
through the Feynman diagrams shown in Fig.~\ref{fig:The axion contribution to muon g2}. The analytical result for each diagram is given by~\cite{Buen-Abad:2021fwq,Ganguly:2022imo}
\begin{align}
    \Delta a_{\mu}^{\rm (a)} = &-\frac{1}{8\pi^2}\Bigg(\frac{C_{ff}}{f_a}\Bigg)^2  m_\mu^2 \int_{0}^1 dx \frac{ m_\mu^2 x^3}{m_a^2(1-x) + m_\mu^2 x^2}, \nonumber\\
    \Delta a_\mu^{\rm (b)} =& \frac{\alpha}{8\pi^3} \left(\frac{C_{ff}}{f_a} \right)^2 m_\mu^2 \Big(\big(I_e + I_\mu + I_\tau + 3I_d + 3I_s + 3I_b  \big) 
    \nonumber\\
    &- \big(3I_u + 3I_c + 3I_t  \big) \Big),
    \label{eq:The analytical expressions for muon g-2}
 \end{align}
 where
 \begin{eqnarray}
     I_{f}(m_a,m_f) = \int_0^{1}dx \frac{m_f^2}{m_f^2 + m_a^2 x(1-x)}\log{\left[ \frac{m_f^2}{m_a^2 x(1-x)}\right]}.
 \end{eqnarray}

The total anomalous magnetic moment of the muon is then given by
\begin{eqnarray}
    \Delta a_{\mu} = \Delta a_{\mu}^a + \Delta a_{\mu}^b.
\end{eqnarray}
It should be noted that diagram~\ref{fig:The axion contribution to muon g2}(a) generates a negative contribution while diagram~\ref{fig:The axion contribution to muon g2}(b) can give either negative or positive ones depending on the axion mass and $C_{ff}$. However, the total muon $a_\mu$ changes sign at the axion mass around 500-600 GeV as shown in Fig.~\ref{fig: total muon g-2}. A plot of $\Delta a_{\mu}$ as a function of axion mass is presented for various values of $C_{ff}=\{100,500,1000,5000,10000\}$ to demonstrate the effect of $C_{ff}/f_a$ on $\Delta a_{\mu}$. Two horizontal lines correspond to the upper and lower limits recently reported by \cite{Aliberti:2025beg,Muong-2:2025xyk}. For $C_{ff}=100$ or $C_{ff}/f_a = 10^{-2}$, we find that the contribution from new diagram has only a mild effect so that the $\Delta a_{\mu}$ is within the range of measurement within $\pm \sigma$ for a range of mass from $\{1,10^4\}$ GeV. For the larger $C_{ff}/f_a$, the smaller allowed range of mass. In addition, the additional $\Delta a_{\mu}$ changes a sign at around 600 GeV. 

\begin{figure}[tbh!]
\centering
\includegraphics[width=\textwidth]{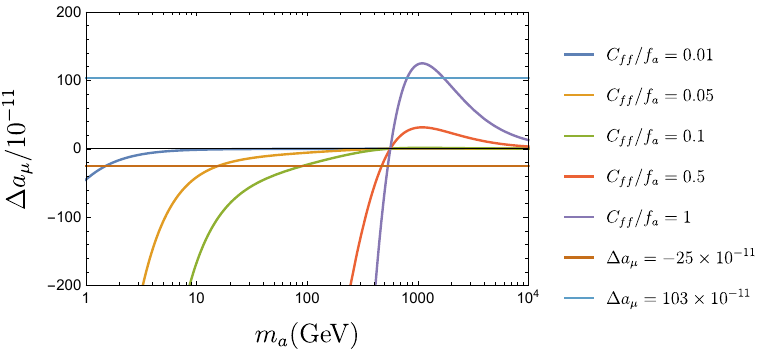}
\caption{The $a_\mu$ plot as a function of $m_a$ for various values of $C_{ff}$ with $f_a = 10^4$ GeV (see legends). Two horizontal lines correspond to upper and lower limits for recent measurement.}
\label{fig: total muon g-2}
\end{figure}

%%%%%%%%%%%%%%%%%%%%%%%%%%%%%%%%%%%%%%%%%
%\section{The results and discussion}
\section{Numerical results}
\label{sec:The results}

In this section, we perform the numerical evaluation of the model parameter space and test our model against the current limits on each sector. The relevant plots and discussion on results are presented.

\subsection*{The neutrino sector}
We first analyze the neutrino sector which depends on seven free parameters. The free parameters are varied within the following ranges:
\begin{align}
    1 \text{ meV }\le\kappa\le1\text{ eV},\quad
    10^{-3}\le y_i,\alpha,\beta\le10^3,\quad
    0\le\phi\le2\pi.
\end{align}
We search for the region of parameter space which is consistent with the neutrino oscillation data at the $3\sigma$ level. We then deduce the corresponding values for $\sum_im_i$, $m_{\nu_e}^{\rm eff}$, and $m_{ee}$.

\begin{table}[htbp]
\centering
\begin{tabular}{|c|c|c|c|c|}\hline
Texture &$\delta_{\rm CP}$ [$^\circ$]& $\sum_im_i$ [meV] &$m^{\mathrm{eff}}_{\nu_e}$ [meV] &$m_{ee}$ [meV]\\
\hline
A &124.16$-$289.95 &58.09$-$59.21 &8.64$-$9.00 &1.16$-$3.04 \\
B &124.18$-$289.93 &58.25$-$59.20& 8.69$-$9.14&1.85$-$4.01 \\
\hline
\end{tabular}
\caption{The range of $\delta_{\rm CP}$, $\sum m_i $, $m_{\nu_e}^{\mathrm{eff}}$ and $m_{ee}$ in each scenario.}
\label{tab:observables}
\end{table}

We find both $m_D^{(A)}$ and $m_D^{(B)}$ can only accommodate NO neutrino masses. As a result, we have $\sum_im_i = \sqrt{\Delta m^2_{\rm sol}} + \sqrt{\Delta m^2_{\rm atm}}$ and 
\begin{equation}
    \left(m_{\nu_e}^{\rm eff}\right)^2 = s_{12}^2c_{13}^2\Delta m^2_{\rm sol} +s_{13}^2\Delta m^2_{\rm atm}.
\end{equation}
Moreover, due to our parametrization of the $M_{\rm light}$, we get $m_{ee} = \kappa$. The ranges of these masses consistent with neutrino oscillation data, are shown in Tab.~\ref{tab:observables}. The values of these effective masses are well below their corresponding experimental bounds.

\begin{figure}[tbh!]
\centering
\begin{subfigure}[b]{0.48\textwidth}
\includegraphics[width=\textwidth]{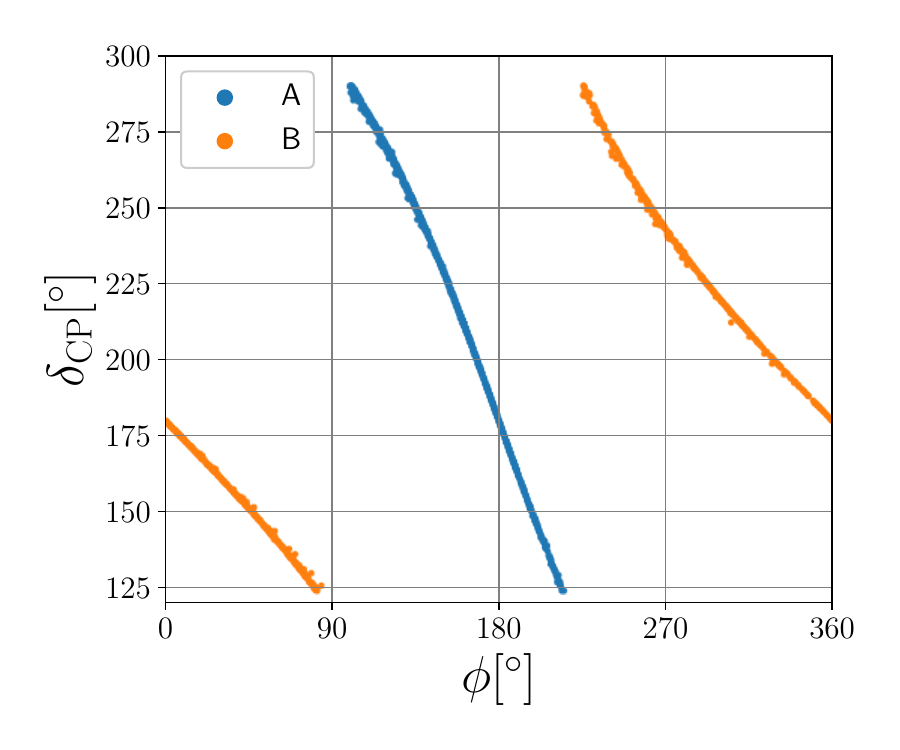}
\centering\small{(a)}
\end{subfigure}
\begin{subfigure}[b]{0.48\textwidth}
\includegraphics[width=\textwidth]{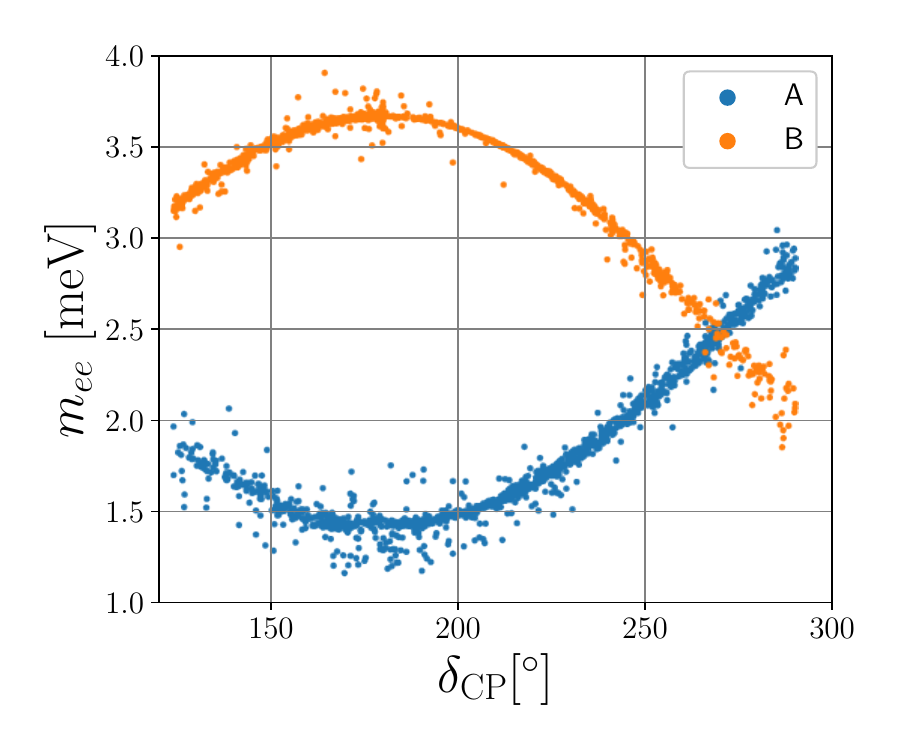}
\centering\small{(b)}
\end{subfigure}
\caption{The relations between the free parameter $\phi$ and $\delta_{\rm CP}$ (a), and $\delta_{\rm CP}$ and $m_{ee}$ (b) in scenarios A and B.}
\label{fig:neutrino plots}
\end{figure}

In our framework, the parameter $\phi$ is responsible for $CP$ violation. We find a strong correlation between $\phi$ and the Dirac phase $\delta_{\rm CP}$ in both scenarios A and B, see Fig.~\ref{fig:neutrino plots}(a). We also observe an upper bound $\delta_{\rm CP}\lesssim290^\circ$ in both scenarios. More interestingly, we have identified a strong correlation between $m_{ee}$ and $\delta_{\rm CP}$, see Fig.~\ref{fig:neutrino plots}(b). Their relationship can be well approximated by an empirical formula
\begin{align}
    m_{ee}^{(\rm A)} &= s_{12}^2c_{13}^2 \sqrt{\Delta m_{\rm sol}^2} + s_{13}^2 \sqrt{\Delta m_{\rm atm}^2} \cos{(\delta_{\rm CP})},\\
    m_{ee}^{(\rm B)} &= s_{12}^2c_{13}^2 \sqrt{\Delta m_{\rm sol}^2} + s_{13}^2 \sqrt{\Delta m_{\rm atm}^2} \cos{(\delta_{\rm CP}+\pi)}.
\end{align}

\begin{figure}[tbh!]
\centering
\begin{subfigure}[b]{0.48\textwidth}
\includegraphics[width=\textwidth]{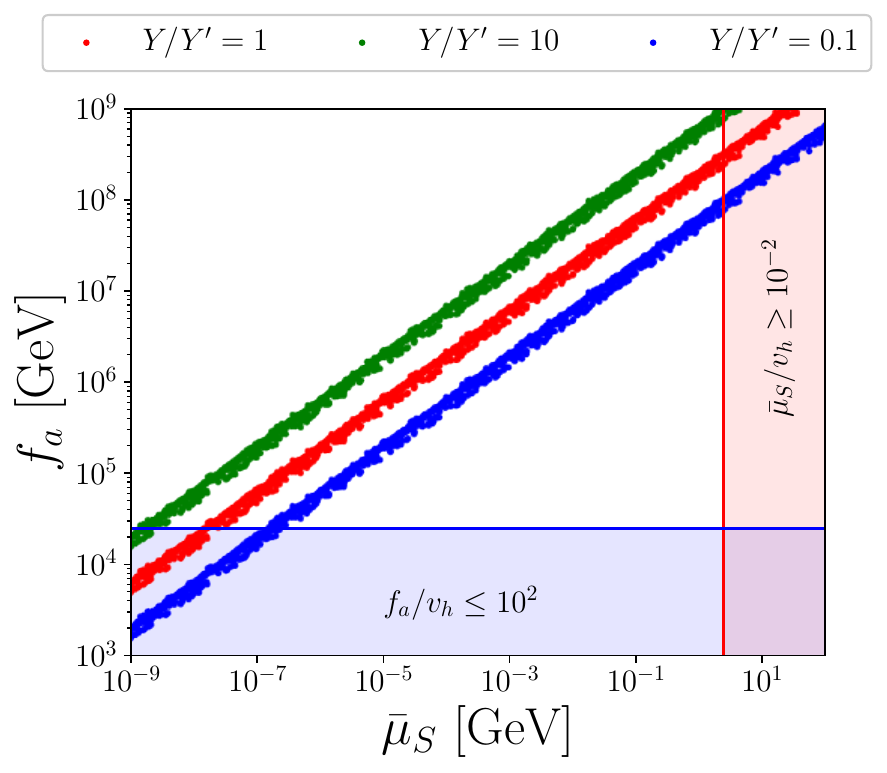}
\centering\small{(a)}
\end{subfigure}
\begin{subfigure}[b]{0.48\textwidth}
\includegraphics[width=\textwidth]{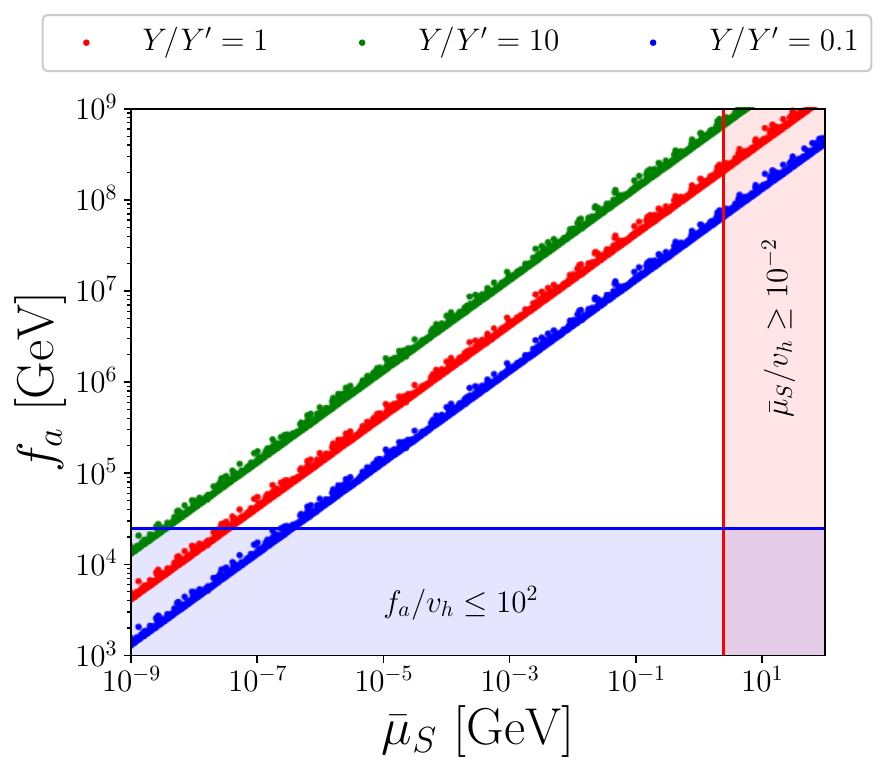}
\centering\small{(b)}
\end{subfigure}
\caption{The allowed region of $\bar \mu_S$ and $f_a$ in scenarios A (left) and B (right). The data is constrained by neutrino experiments. The shaded regions indicates where the approximation formula of neutrino mass  Eq.~(\ref{eq: light neutrino mass}) does not hold.}
\label{fig: muS and fa plots}
\end{figure}

% \begin{figure}[tbh!]
% \centering
% \begin{subfigure}[b]{0.48\textwidth}
% \includegraphics[width=\textwidth]{fig/BR_A_hist.pdf}
% \centering\small{(a)}
% \end{subfigure}
% \begin{subfigure}[b]{0.48\textwidth}
% \includegraphics[width=\textwidth]{fig/BR_B_hist.pdf}
% \centering\small{(b)}
% \end{subfigure}
% \caption{The distribution of ${\rm Br}(\mu \to e \gamma)$ from 2018 sample points that obey the neutrino oscillation data.}
% \label{fig:mu to e gamma hist}
% \end{figure}

\begin{figure}[tbh!]
\centering
\begin{subfigure}[b]{0.48\textwidth}
\includegraphics[width=\textwidth]{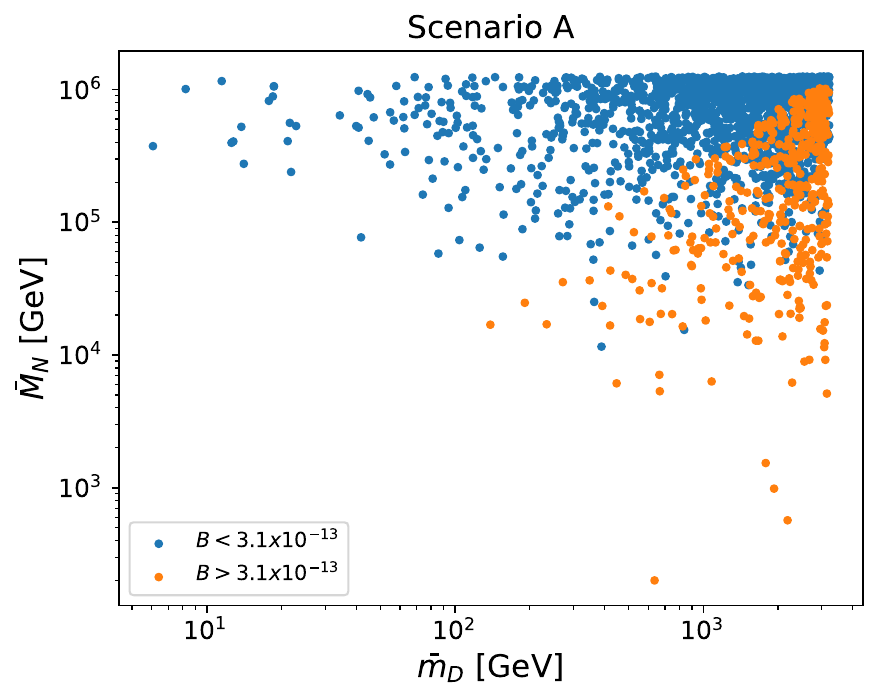}
\centering\small{(a)}
\end{subfigure}
\begin{subfigure}[b]{0.48\textwidth}
\includegraphics[width=\textwidth]{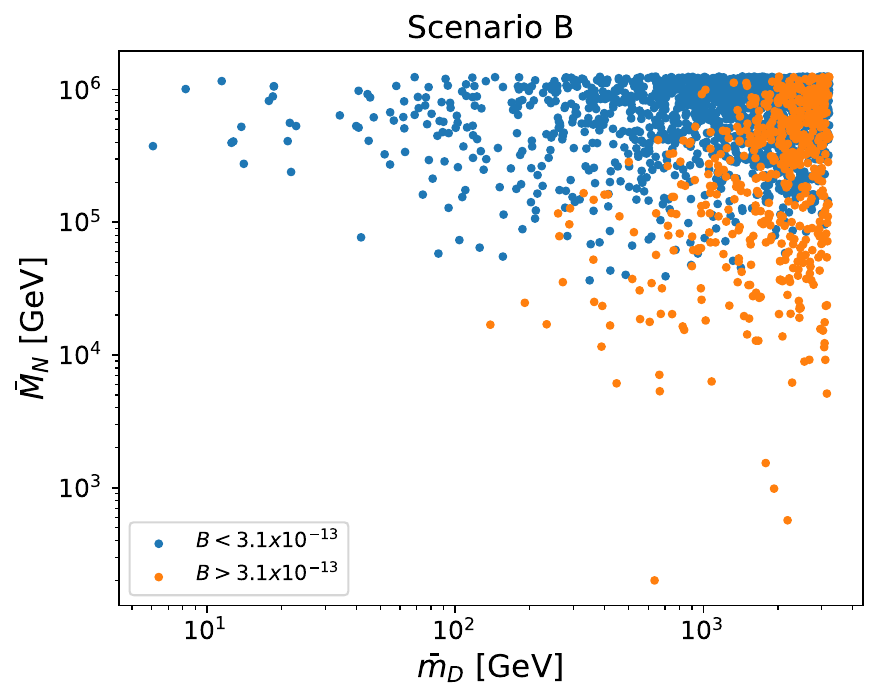}
\centering\small{(b)}
\end{subfigure}
\caption{The data set from neutrino oscillation validated by cFLV constraint ${\rm B}(\mu^+ \to e^+\gamma)_{\rm MEG\;II}<3.1 \times 10^{-13}$. }
\label{fig: mu e gamma plots}
\end{figure}

By using the data from multi-dimensional scan constrained by neutrino data and projecting on the $\bar \mu_S$ and $f_a$ plane, the allowed region is presented in Fig.~\ref{fig: muS and fa plots} (a) and (b) for scenario A and B, respectively. Here the shaded regions represent parameter space where the approximate formula for neutrino mass Eq.~(\ref{eq: light neutrino mass}) is no longer valid. The red, green blue points indicate data with $Y/Y^\prime = 1,10$ and $0.1$, respectively. While an obvious correlation is expected from Eq.~(\ref{eq: kappa}), the deviation results from the propagated uncertainties in neutrino observables.

In addition to constraining the parameter space of neutrino sector by using neutrino oscillation data, we also find that these points are allowed by the cLFV constraint where the most stringent limit is given by MEG II ${\rm B}(\mu^+ \to e^+ \gamma) < 3.1 \times 10^{-13}$~\cite{MEGII:2025gzr}. We use the data set which is allowed by neutrino oscillation data. Then we make a random scan on the parameter $\bar M_N$ and $\bar m_D$ to compute the branching ratio of $\mu^+ \to e^+ \gamma$. In Fig.~\ref{fig: mu e gamma plots}, the blue and orange points are allowed and disallowed by MEG II data, respectively. Since the ${\rm B}(\mu^+ \to e^+ \gamma) \propto \sum_i W_{\mu i}^*W_{e i} \propto \bar m_D^2/\bar M_N^2$, so the point with a large ratio of $\bar m_D/\bar M_N$ is like to be excluded. 

\begin{figure}[t]
\centering
% \begin{subfigure}[b]{0.48\textwidth}
% \includegraphics[width=\textwidth]
% {fig/plot_mchi_caafa_fa1e4_Xphi1.pdf}
% \centering\small{(a)}
% \end{subfigure}
% \begin{subfigure}[b]{0.48\textwidth}
% \includegraphics[width=\textwidth]
% {fig/plot_mchi_caafa_fa1e5_Xphi1.pdf}
% \centering\small{(b)}
% \end{subfigure}
\begin{subfigure}[b]{0.48\textwidth}
\includegraphics[width=\textwidth]
{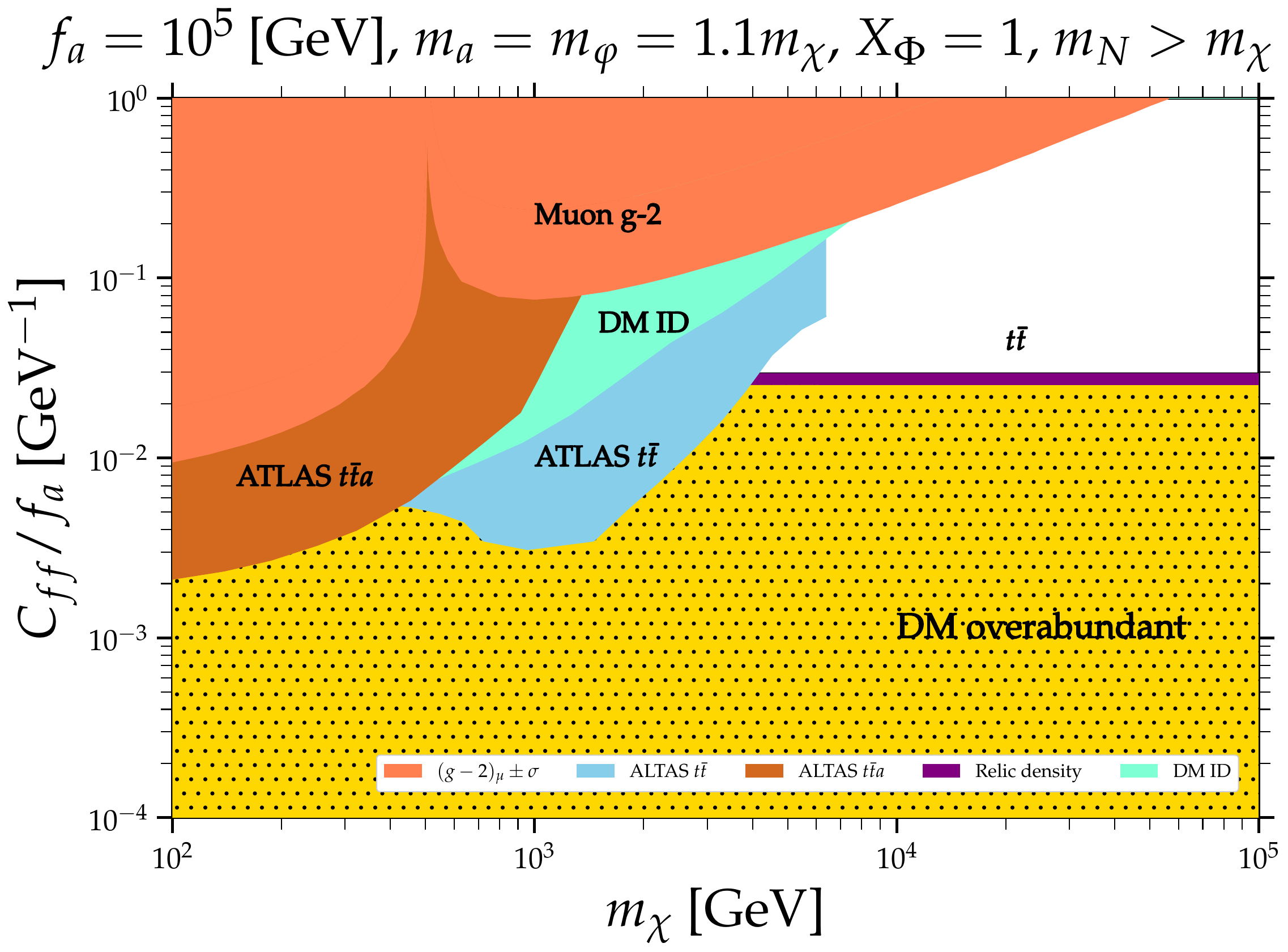}
\centering\small{(a)}
\end{subfigure}
\begin{subfigure}[b]{0.48\textwidth}
\includegraphics[width=\textwidth]
{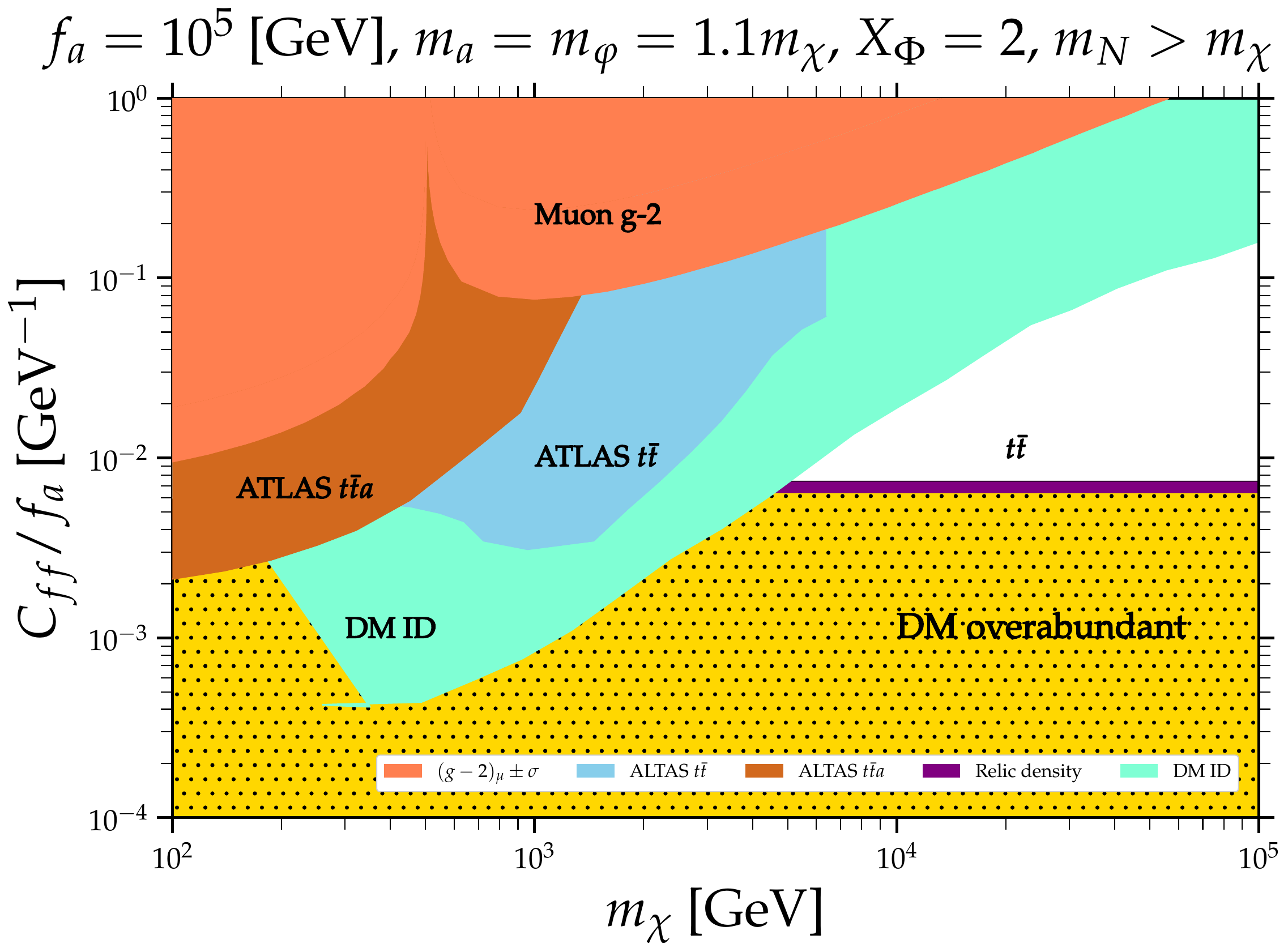}
\centering\small{(b)}
\end{subfigure}
\caption{The plot of $C_{ff}/fa$ [GeV]$^{-1}$ and $m_{\chi}$ [GeV] for $m_a = m_\varphi = 1.1 m_\chi$ and $m_{\nu_{\rm heavy}}> m_{\chi}$. The orange colour labels the region excluded by latest $g-2$ within $\pm\sigma$. The blue and brown regions represent the exclusion limit from the ATLAS measurement of $t\bar{t}$ and $t\bar{t}a$ production, respectively. The green region is combined exclusion limit from the Fermi-LAT, MAGIC, VERITAS, HAWG and HESS experiment. The purple region is consistent with by thermal relic density. The white and dotted gold represent the under and over abundant regions, respectively. A pair of $f\bar f$ label the dominant DM annihilation channel. Note that constraints from DM direct detection are much weaker than the constraints shown here, and hence are omitted.
%\textcolor{blue}{ The current limit on DM direct detection (DM DD) from LZ(2025) cannot be shown in the range of the presented plot as it is far above after constraints.}
}
\label{fig: plot of Cff and mchi}
\end{figure}

\subsection*{The DM sector}

For the DM and muon $g-2$ sector, we consider five free parameters: the DM mass ($m_\chi$), the axion mass ($m_a$), the axion decay constant ($f_a$), the ratio of $f_a/v_\varphi$ ($X_{\Phi}$), and the axion-fermion coupling coefficient ($C_{ff}$). We aim to identify the parameter space consistent with three key observations: 1) the observed DM relic abundance, 2) the measured muon magnetic moment, and 3) the axion-$t\bar t$ coupling search conducted by ATLAS.

Fig.~\ref{fig: plot of Cff and mchi} illustrates the allowed and excluded regions across various current constraints for $m_{\chi}>200$ GeV. In this range of DM mass, the dominant channel of DM annihilation is $\chi \bar\chi \to t \bar t$. The orange region is excluded by recent muon $g-2$ measurements. The purple region represents the parameter space consistent with the thermal relic density, assuming the freeze-out approximation. The blue and brown regions indicate the exclusion limit derived from the ATLAS measurement of $t\bar{t}$ and $t\bar{t}a$ production, respectively, which is used to constrain axion-top coupling. This measurement, detailed in Ref. \cite{Esser:2023fdo}, considered the lepton+jets channel with high-$p_T$ top quarks and an integrated luminosity of $\mathcal{L} = 139$ fb$^{-1}$. The green region is combined exclusion limit from the Fermi-LAT, MAGIC, VERITAS, HAWG and HESS experiments~\cite{Fermi-LAT:2025gei}, which will be referred to as the dark matter indirect detection limit (DM ID). The white region indicates under-abundant DM, while dotted gold signifies over-abundant regions. In addition, the current DM direct detection (DM DD) limit is represented by LZ experiments measuring the spin-dependent cross section of DM scattering on protons and neutrons~\cite{LZ:2024zvo} which would require values of $C_{ff}/f_a$ far above the range shown here, and hence lies outside the plot. The separate plots of $\sigma_{\rm SD}^{p}$ and $\sigma_{\rm SD}^{n}$ as a function of DM mass are presented in Fig. \ref{fig: direct detection limit}.

Fig.~\ref{fig: plot of Cff and mchi} (a) and (b) demonstrate that increasing $X_\Phi$ necessitates a corresponding decrease in $C_{ff}$. This is because the factor $C_{ff}^2 X_\Phi^4$ must remain constant for specific values of $m_a$ and $m_\varphi$ to maintain the relic density. Fig.~\ref{fig: plot of Cff and mchi} (a), the most significant constraint comes from the ATLAS $t\bar{t}$ search, which excludes DM masses $4$ TeV. When $X_{\Phi}$ is doubled, as shown in Fig. \ref{fig: plot of Cff and mchi} (b), the factor $C_{ff}$ must be suppressed by a factor of four, given that the annihilation cross-section scales as $\braket{\sigma v} \propto C_{ff}^2 X_{\Phi}^4$. The most important limit is therefore given by the DM ID which exclude DM mass below 5 TeV. 

In Fig.~\ref{fig: plot of Cff and mchi}, we assume that the heavy neutrino masses are larger than the DM mass such that they do not play any role in evaluating the DM relic density. However, if they were lighter than DM candidate, they would contribute to the relic density. For example, if they were heavier than top quark, $m_{t}<m_{\nu_{\rm heavy}}<m_{\chi}$, their contributions would appear as a lower step in the $C_{ff}/f_a$ versus $m_{\chi}$ plot, allowing for a lower region of $C_{ff}/f_a$.

\begin{figure}[thb!]
\centering
\begin{subfigure}[b]{0.48\textwidth}
\includegraphics[width=\textwidth]
{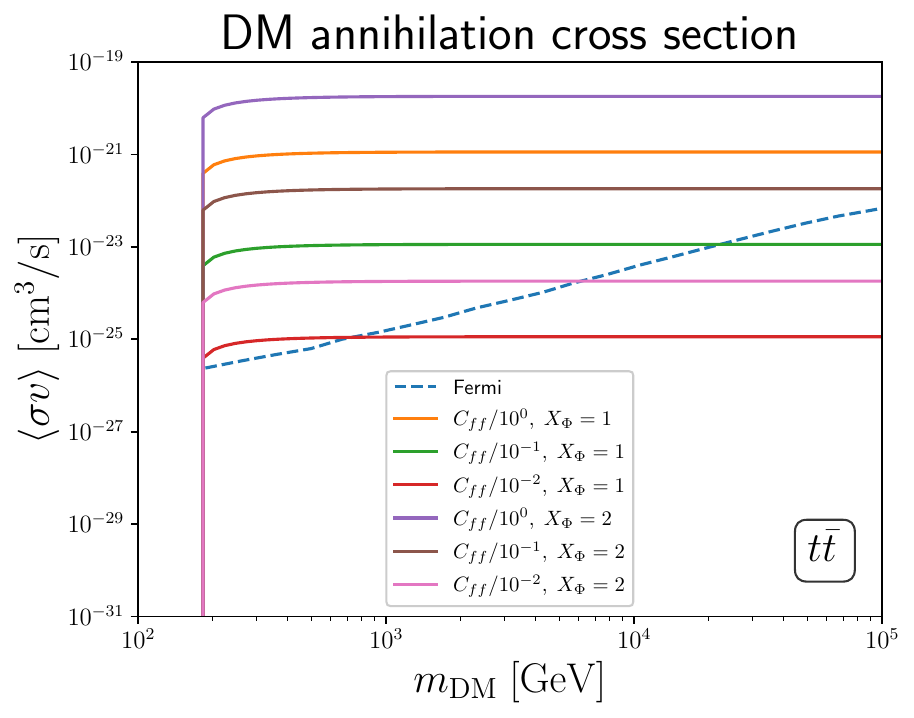}
\centering\small{(a)}
\end{subfigure}
% \begin{subfigure}[b]{0.48\textwidth}
% \includegraphics[width=\textwidth]
% {fig/plot_bb.pdf}
% \centering\small{(b)}
% \end{subfigure}
% \begin{subfigure}[b]{0.48\textwidth}
% \includegraphics[width=\textwidth]
% {fig/plot_ll.pdf}
% \centering\small{(c)}
% \end{subfigure}
% \begin{subfigure}[b]{0.48\textwidth}
% \includegraphics[width=\textwidth]
% {fig/plot_mm.pdf}
% \centering\small{(d)}
% \end{subfigure}
% \begin{subfigure}[b]{0.48\textwidth}
% \includegraphics[width=\textwidth]
% {fig/plot_ee.pdf}
% \centering\small{(e)}
% \end{subfigure}
\caption{The indirect detection limit on DM annihilation is dominated by the process $\chi \bar \chi \to t \bar t$.}
\label{fig: indirect detection limit}
\end{figure}

In addition to the DM relic density, the indirect detection constraint is also crucial to the parameter space, especially, the $t\bar t$ channel which is the dominant one. In Fig.~\ref{fig: indirect detection limit}, all parameter space with $C_{ff}/f_a = 1$ [GeV]$^{-1}$ are excluded (purple, orange and brown lines). The green line with $C_{ff}/f_a = 0.1$ [GeV]$^{-1}$ is allowed for $m_{\rm DM}> 2$ TeV while the red and pink lines are allowed for $m_{\rm DM}> 1$ TeV and 6 TeV, respectively. This limit is consistent with exclusion limits in Fig.~\ref{fig: plot of Cff and mchi} in which the parameter space is allowed for $m_{\rm DM}>6$ TeV and $C_{ff}/f_a = 0.1$ [GeV]$^{-1}$.  

\begin{figure}[tbh!]
\centering
\begin{subfigure}[b]{0.48\textwidth}
\includegraphics[width=\textwidth]
{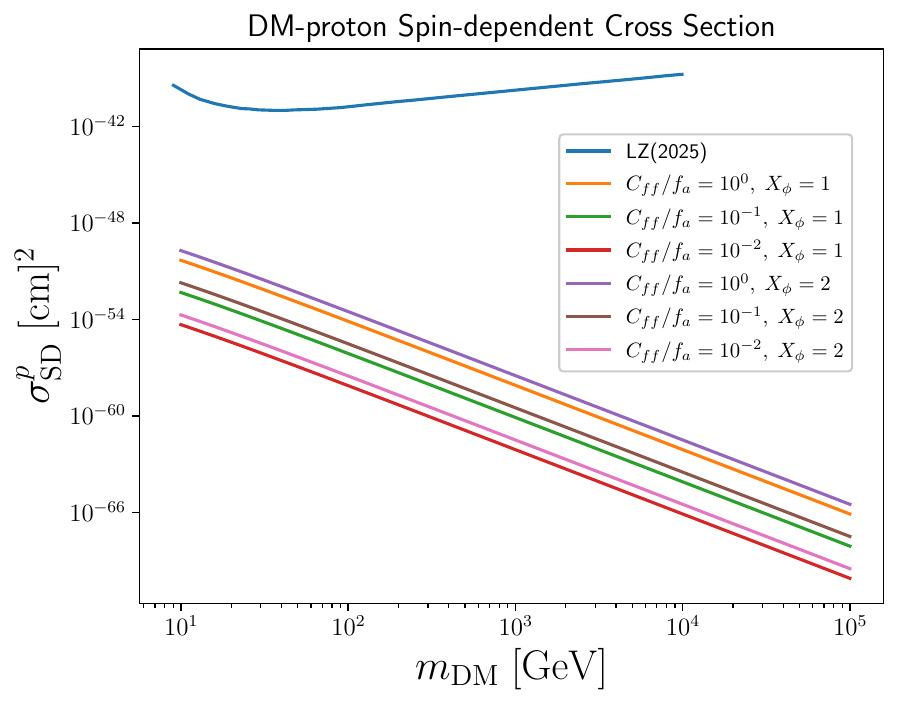}
\centering\small{(a) DM-Proton scattering}
\end{subfigure}
\begin{subfigure}[b]{0.48\textwidth}
\includegraphics[width=\textwidth]
{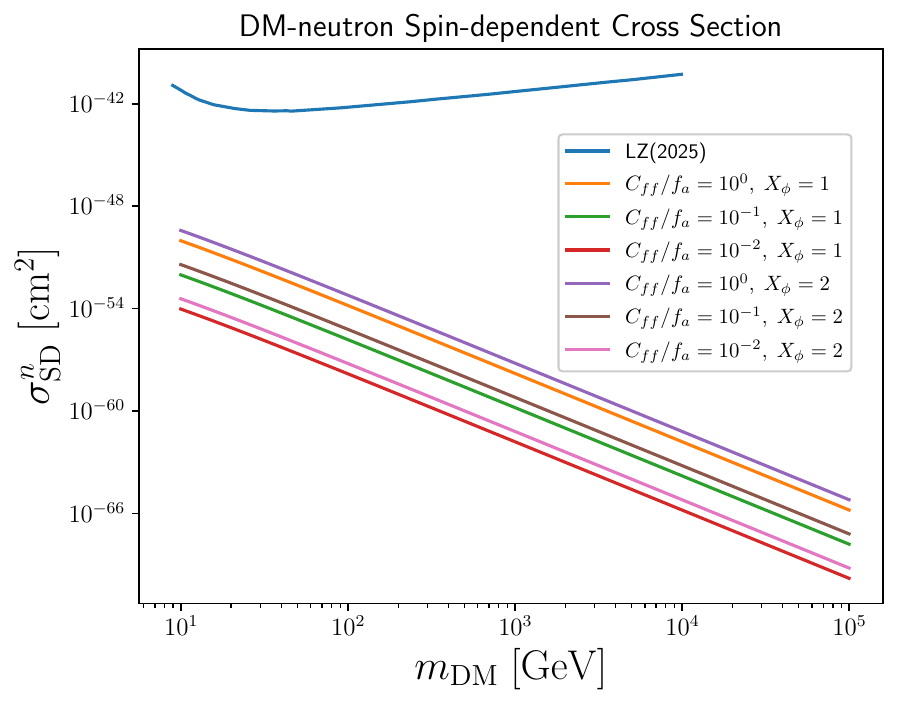}
\centering\small{(b) DM-neutron scattering}
\end{subfigure}
\caption{The direct detection limits on DM elastic scattering cross section into relevant SM channels.}
\label{fig: direct detection limit}
\end{figure}

In our model, the DM-nucleon scattering is spin-dependent due to the pseudoscalar interaction of axion and other fermions (SM ones and DM). In Fig.~\ref{fig: direct detection limit} we calculate the spin-dependent scattering cross section of DM-proton and DM-neutron and compare them with the current constraint from LZ collaboration. We find that the predicted cross section is far below the present limit such that it cannot be used to exclude our parameter space and is not relevant here.

%resulting in the lower region of $C_{ff}/f_a$ is allowed.

% \begin{figure}[t]
% \centering
% %\includegraphics[width=\textwidth]{fig/AxionPhoton_ColliderBounds.pdf}
% \begin{subfigure}[b]{0.7\textwidth}\includegraphics[width=\textwidth]{fig/AxionPhoton_ColliderBounds_fa1e4.pdf}
% %\centering\small{(a)}
% \end{subfigure}\\
% \begin{subfigure}[b]{0.7\textwidth}\includegraphics[width=\textwidth]{fig/AxionPhoton_ColliderBounds_fa1e5.pdf}
% %\centering\small{(b)}
% \end{subfigure}
% \caption{The region of parameter space on the $m_a-C_{\gamma\gamma}/f_a$ plane consistent with the DM relic density ({purple}) and can explain the muon \textcolor{red}{$g-2$ anomaly} ({green, teal and deep blue} for $C_{\mu\mu}/f_a = -0.01$, $-0.1$ and $-1$ [GeV$^{-1}$] respectively). In these plots, $m_a=m_\varphi=1.1m_\chi$. The hatched region is ruled out by the gamma ray line search and the dotted region produces an over abundance of DM. In the top (bottom) figure, $f_a$ = 10 (100) TeV.}
% \label{fig: all collider constraints on ALP}
% \end{figure}

\begin{figure}[tbh!]
\centering
\begin{subfigure}[b]{0.48\textwidth}\includegraphics[width=\textwidth]{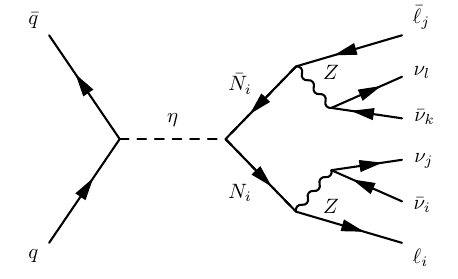}
\centering\small{(a) $pp\to \ell_i \bar{\ell}_j + \slashed{E}_T$}
\end{subfigure}
\begin{subfigure}[b]{0.48\textwidth}\includegraphics[width=\textwidth]{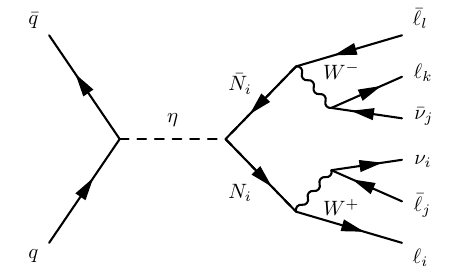}
\centering\small{(b)  $pp\to \ell_i \bar{\ell}_j \ell_k \bar{\ell}_l + \slashed{E}_T$}
\end{subfigure}
\begin{subfigure}[b]{0.48\textwidth}\includegraphics[width=\textwidth]{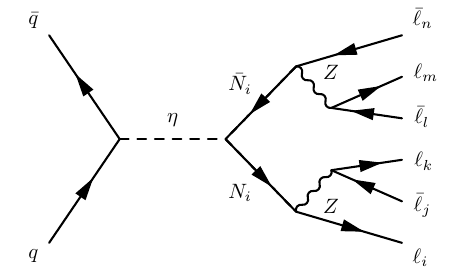}
\centering\small{(c)  $pp\to \ell_i \bar{\ell}_j \ell_k \bar{\ell}_l \ell_m \bar{\ell}_n$}
\end{subfigure}
\caption{The possible signatures of heavy neutrino searches at LHC or FCC-hh consisted of a) the dileptons channel, b) the 4-lepton channel and c) the 6-lepton channel. These search channels can also be used to probe the new scalar particles like $\eta=\{a,\varphi\}$.}
\label{fig: LHC signatures for heavy neutrino searches}
\end{figure}

%%%%%%%%%%%%%%%%%%%%%%%%%%%%%%%%%%%%%%%%%
\section{Conclusion and discussion}
\label{sec:Conclusion}

In this work, we proposed a unified framework where the spontaneous breaking of a single global $U(1)_{\rm PQ}$ symmetry simultaneously generates the mass scales for the inverse seesaw mechanism and stabilizes a fermionic DM candidate.
%We presented a minimal extension of the SM that combines the inverse seesaw mechanism with an ALP portal DM. This model addresses both the origin of neutrino masses and the nature of DM while remaining consistent with current experimental constraints. 
Unlike traditional seesaw models that require high mass scales, the inverse seesaw mechanism allows for TeV-scale new physics, making it potentially accessible in future collider experiments. Additionally, the ALP coming from the $U(1)_{\rm PQ}$ breaking serves as a mediator between the SM and DM sectors, providing a distinct phenomenology that includes a positive contribution to the muon anomalous magnetic moment (\(g-2\)) and compatibility with observed relic density.

From our analysis, we find two minimal scenarios for the inverse seesaw mechanism. They both lead to NO neutrinos with vanishing lightest neutrino mass. As a result, the total neutrino mass $\sum_i m_i$ and the effective mass $m_{\nu_e}^{\rm eff}$ and $m_{ee}$ are directly related to $\Delta m^2_{\rm sol}$ and $\Delta m^2_{\rm atm}$. Moreover, we found a strong correlation between $m_{ee}$ and $\delta_{\rm CP}$ which makes our model more distinctive than others. For the total neutrino mass, we find $\sum_i m_i\simeq 58$ meV. This is an order of magnitude below the current Planck constraint. However, it can be probed by the Simon Observatory whose projected sensitivity is $\sum_im_i\le40$ meV at 95\% CL~\cite{SimonsObservatory:2019qwx}. Meanwhile, the effective mass $m_{\nu_e}^{\rm eff}$ in our scenarios is around 9 meV, which is an order of magnitude below the projection $m_{\nu_e}^{\rm eff}\le200$ meV at 90\% CL of both the HOLMES and the KATRIN experiments~\cite{Alpert:2014lfa,KATRIN:2019yun}. Moreover, we find the effective mass $m_{ee}$ is of order a few meV. This is well below the projected sensitivity of $m_{ee}\le 13-29$ meV at 90\% CL of the LEGEND experiment phase II~\cite{LEGEND:2017cdu}. Thus, the key signature for our neutrino sector is a signal in cosmological measurements ($\sum_im_i$) without corresponding signals in laboratory based measurements ($m_{\nu_e}^{eff}$ and $m_{ee}$).

In our scenario, the ALP-gauge boson coupling is only generated at the loop level and contribute less than 1\% compared to the tree level one. Therefore, the DM density is purely determined from the annihilation $\chi\bar\chi\to f \bar f$ 
through an s-channel ALP exchange. For sufficiently large $m_\chi$, the $t\bar t$ annihilation channel dominates over other di-fermion channels. In the benchmark scenario where $m_a=m_\varphi=1.1m_\chi$, we find that DM can be a thermal relic for $m_a\gtrsim 4$ TeV, see Fig.~\ref{fig: plot of Cff and mchi}. The ALP couplings that determine DM density also generate either a positive or a negative shift in the muon $g-2$. We identify the region of parameter space in the $m_a-C_{ff}/f_a$ plane which could explain the DM relic density while satifying the DM ID, the muon $g-2$, and the ATLAS $t\bar t$ constraints for $f_a/[{\rm GeV}] = 10^5$ and $X_{\Phi}=1,2$, respectively. 

Finally, the heavy neutrinos in our model could potentially be probed by the LHC and FCC-hh through the process of the proton-proton collision producing the scalar particles $\{h,a,\varphi\}$ which then decay into a pair of heavy neutrinos. These heavy neutrinos will eventually decay into a product of di-leptons, four-leptons and six-leptons, i.e.,
$pp \to a,h,\varphi \to N_i \bar{N}_j \to \ell_i \bar{\ell}_j + \slashed{E}_T,\; \ell_i \bar{\ell}_j \ell_k \bar{\ell}_l + \slashed{E}_T,\; \ell_i \bar{\ell}_j \ell_k \bar{\ell}_l \ell_m \bar{\ell}_n$, see Fig. \ref{fig: LHC signatures for heavy neutrino searches}. Notice that these processes can be utilised in searches for new heavy scalars, such as an axion or $\varphi$, through the resonances in the dilepton invariant mass. We leave the study of the collider phenomenology for future work.  

% \textcolor{blue}{In addition, the dark matter direct detection rate is very small compared to the experimental bound due to the spin-dependent suppression of the axion-nucleon interaction}. If neutrino and DM masses are generated through our scenario, the LHC Run 3 and/or HL-LHC data could probe the $m_{\chi}\lesssim 2$ TeV regime for $f_a = 10^4$ GeV scenario. \textcolor{red}{CP to PU: Since we already have excluded $m_{\chi}\lesssim 2$ TeV in our new analysis, should this be updated according to the HL-LHC projection as well?}

\section*{Acknowledgement}
This research has received funding support from the NSRF via the Program
Management Unit for Human Resources \& Institutional Development, Research and
Innovation [grant number B13F670063]. CP is supported by Fundamental Fund 2567 of Khon Kaen University via the National Science Research and Innovation Fund (NSRF).
The work of PU is supported in part by the Mid-Career Research Grant from the National Research Council of Thailand under contract no. N42A650378. PU also thanks the High-Energy Physics Research Unit, Chulalongkorn University for hospitality while part of this work was being completed.

\appendix

\section{The neutrino mixing matrix}
\label{sec: full neutrino mixing matrix}

The process of diagonalization of mass matrix defined in Eq.(\ref{eq: the mass matrix of the model}) is the following.
First, we rotate the mass matrix into a diagonal block matrix
\begin{align}
    V^T 
    \begin{pmatrix}
        \mathbf{0} & \mathcal{M}_D \\
        \mathcal{M}_D^T & \mathcal{M}_S
    \end{pmatrix} V 
    = \begin{pmatrix}
        M_{\rm light} & \bf0\\
        \bf 0 & M_{\rm heavy}  
    \end{pmatrix},
\end{align}
where $M_{\rm light}$ and $M_{\rm heavy}$ are defined by (\ref{eq: light neutrino mass}) and (\ref{eq: heavy neutrino mass}), respectively. However, $M_{\rm light}$ and $M_{\rm heavy}$ are not yet diagonal. The rotation matrix of the first diagonalization reads
\begin{align}
    V = \begin{pmatrix}
        \mathbf{1}_3 -\frac{1}{2}\mathcal{M}^*_D(\mathcal{M}_S^*)^{-1}(\mathcal{M}_S^T)^{-1}\mathcal{M}_D^T & \mathcal{M}^*_D(\mathcal{M}_S^*)^{-1} \\
        - (\mathcal{M}_S^T)^{-1}\mathcal{M}_D^T & \mathbf{1}_6 -\frac{1}{2}(\mathcal{M}_S^T)^{-1}\mathcal{M}_D^T \mathcal{M}^*_D(\mathcal{M}_S^*)^{-1}
    \end{pmatrix} + \mathcal{O}(\xi^3)
\end{align}
where $\xi \equiv \mathcal{M}_D(\mathcal{M}_S)^{-1}$ and $\mathcal{M}_{D}$ and $\mathcal{M}_{S}$ are defined as
\begin{eqnarray}
    \mathcal{M}_D &=& \begin{pmatrix}
        m_D \;\;\;& 0
    \end{pmatrix},\\
    \mathcal{M}_S &=& \begin{pmatrix}
     \mu_R & M^T_N \\
     M_N & \mu_S
    \end{pmatrix}.
\end{eqnarray}
The second rotation is simply perform by using 
\begin{align}
    U = \begin{pmatrix}
        U_{\rm PMNS} & \bf 0\\
        \bf 0 & U_{\rm R} 
    \end{pmatrix}
    \label{eq: definition of rotation matrix U}
\end{align} such that
\begin{align}
    U^T \begin{pmatrix}
        M_{\rm light} & \bf 0\\
        \bf 0 & M_{\rm heavy} 
    \end{pmatrix}
    U
    =
    \begin{pmatrix}
        m_\nu & \bf 0\\
        \bf 0 & m_R
    \end{pmatrix}
\end{align}
where $m_\nu = {\rm diag}.(m_1,m_2,m_3)$ and $m_R = {\rm diag.}(m_4,m_5,m_6,m_{7})$. This matrix equation can be split into two equations
\begin{align}
    m_\nu &= U_{\rm PMNS}^T M_{\rm light} U_{\rm PMNS},\nonumber\\
    m_R &= U_{\rm R}^T M_{\rm heavy} U_{\rm R}.
    \label{eq: expression for mnu and mR}
\end{align}
The PMNS rotation matrix can be parameterised by three mixing angles, one Dirac and two Majorana CP phases which is given by
\begin{align}
    U_{\rm PMNS} =
    \begin{pmatrix}
        c_{12}c_{13} & s_{12}c_{13} & s_{13}e^{-i\delta} \\
        -s_{12}c_{23}-c_{12}s_{23}s_{13}e^{i\delta} & c_{12}c_{23} -s_{12}s_{23}s_{13}e^{i\delta} & s_{23}c_{13}\\
        s_{12}s_{23} - c_{12}c_{23}s_{13}e^{i\delta} & -c_{12}s_{23}-s_{12}c_{23}s_{13}e^{i\delta} & c_{23}c_{13}
    \end{pmatrix}
    \begin{pmatrix}
        e^{i\alpha_1/2} & 0& 0\\
        0 & e^{i\alpha_2/2} & 0 \\
        0 & 0 &1
    \end{pmatrix}
\end{align}
where we use the shorthand for $s_{ij}\equiv \sin{\theta_{ij}}$ and $c_{ij}\equiv \cos{\theta_{ij}}$. 

Combining two rotation matrices into a single matrix is straightforward. We define $W = VU$ where
\begin{align}
   W \simeq \begin{pmatrix}
        \left[\mathbf{1}_3 -\frac{1}{2}\mathcal{M}^*_D(\mathcal{M}_S^*)^{-1}(\mathcal{M}_S^T)^{-1}\mathcal{M}_D^T\right]U_{\rm PMNS} & \mathcal{M}^*_D(\mathcal{M}_S^*)^{-1}U_{\rm R} \\
        - (\mathcal{M}_S^T)^{-1}\mathcal{M}_D^T U_{\rm PMNS} & \left[\mathbf{1}_6 -\frac{1}{2}(\mathcal{M}_S^T)^{-1}\mathcal{M}_D^T \mathcal{M}^*_D(\mathcal{M}_S^*)^{-1}\right] U_{\rm R},
    \end{pmatrix}.
\end{align}
such that
\begin{align}
    W^T
    \begin{pmatrix}
        \bf 0 & \mathcal{M}_D \\
         \mathcal{M}_D^T & \mathcal{M}_S
    \end{pmatrix}
    W = \begin{pmatrix}
        m_\nu & \bf 0\\
        \bf 0 & m_R
    \end{pmatrix}.
\end{align}

Let us express the interaction and mass eigenstates of the neutrino mass matrix as $\psi_{R,\rm int} = (\nu_{\alpha L}^c, N_{iR}, S_{jL}^c)^T$ and $\psi_{R,\rm mass} = (n_{1R},n_{2R},n_{3R},N_{1R},N_{2R},N_{3R},N_{4R})^{T}$ such that $\psi_{R,\rm int} = W \psi_{R,\rm mass}$ where
\begin{align}
    \vec{\nu}_{L}^c &=  \bigg(\left[\mathbf{1}_3 -\frac{1}{2}\mathcal{M}^*_D(\mathcal{M}_S^*)^{-1}(\mathcal{M}_S^T)^{-1}\mathcal{M}_D^T\right]U_{\rm PMNS}\bigg)\vec{n}_R + \bigg(\mathcal{M}^*_D(\mathcal{M}_S^*)^{-1}U_{\rm R}\bigg)\vec{\mathcal{N}}_R,\nonumber\\ 
    \vec{\mathcal{S}}_{R}&= - \bigg((\mathcal{M}_S^T)^{-1}\mathcal{M}_D^T U_{\rm PMNS}\bigg)\vec{n}_R  + 
    \bigg(\left[\mathbf{1}_6 -\frac{1}{2}(\mathcal{M}_S^T)^{-1}\mathcal{M}_D^T \mathcal{M}^*_D(\mathcal{M}_S^*)^{-1}\right] U_{\rm R}\bigg)\vec{\mathcal{N}}_R.
    \label{eq: the physical states of neutrinos}
\end{align}
Here $\vec \nu_{L}^c = (\nu_e^c,\nu_\mu^c,\nu_\tau^c)^T$, $\mathcal{\vec S}_{R} = (N_{1R},N_{2R},S_{1L}^c,S_{2L}^c)^T$, $\vec n_{R} = (n_{1R},n_{2R},n_{3R})^T$ and $\mathcal{\vec N}_{R} = (N_{1R},N_{2R},N_{3R},N_{4R})^T$

\section{The branching ratio for Lepton Flavour Violation}
\label{sec: the mu to e gamma rate}

    The generic branching ratio for for $\ell \to \ell^\prime \gamma$ is given by \cite{DiMeglio:2024gve,Ilakovac:1994kj} 
    \begin{eqnarray}
        B(\ell \to \ell^\prime \gamma) = \frac{\alpha^3_w s_w^2}{256 \pi^2}\frac{m_\ell^4}{M_W^4} \frac{m_{\ell}}{\Gamma_\ell}|G_{\gamma}^{\ell \ell^\prime}|^2,
        \label{eq: cLFV rate}
    \end{eqnarray}
    where $\alpha_w = g_w^2/4\pi$, $\Gamma_\ell$ is the total decay width of the charged lepton $\ell$ and $G^{\ell\ell^\prime}_{\gamma}$ is the loop function.
    \begin{eqnarray}
        G_{\gamma}^{\ell\ell^\prime} = W_{\ell i}^* W_{\ell^\prime i} G_{\gamma},
    \end{eqnarray}
    and
    \begin{eqnarray}
        G_{\gamma}(x) = \left\{ 
        \begin{array}{ll}
           -\frac{2x^3 + 5x^2 - x}{4(1-x)^3} -\frac{3x^3}{2(1-x)^4}\ln{x} &;  x\ll 1\\
           \frac{1}{2}  &;  x\gg 1.
        \end{array}
        \right.
    \end{eqnarray}
The mixing matrices between charged lepton can intermediate state is given by $W_{\ell i}$. In the case of $\mu \to e \gamma$, the charged-Lepton-Flavour-Violation becomes
\begin{eqnarray}
        B(\mu \to e \gamma) = \frac{\alpha^3_w s_w^2}{256 \pi^2}\frac{m_\mu^4}{M_W^4} \frac{m_{\mu}}{\Gamma_\mu}|G_{\gamma}^{\mu e}|^2,
        \label{eq: BR of mu to e gamma}
    \end{eqnarray}
where 
\begin{eqnarray}
    G^{\mu e}_{\gamma} = \sum_{i=1,2,...,7} W_{\mu i}^* W_{e i} G_\gamma\left(\frac{m_i^2}{M_W^2}\right).
    \label{eq: mixing matrix for BR}
\end{eqnarray}
We use both Eq.(\ref{eq: BR of mu to e gamma}) and (\ref{eq: mixing matrix for BR}) to calculate the branching ratio of $\mu \to e \gamma$.

\section{The cross section of $\chi\bar\chi \to \gamma\gamma$}
\label{sec: a triangular loop function}

We can use the axion-photon coupling discussed in section~\ref{sec:the axion-fermion interaction} to compute the annihilation cross section of DM into diphoton 
\begin{eqnarray}
    i \mathcal{M} =&& \left(-i \frac{m_\chi}{v_\varphi}\right) \bar v(p_2,s) \gamma_5 u(p_1,s^\prime) \frac{i}{(p_1 + p_2)^2 - m_a^2} \nonumber\\
    && \cdot \sum_f \frac{C_{ff}}{f_a} 2 \log{(2)^2} \alpha_{\rm EM} \left(\frac{m_f^2}{s}\right) \epsilon^{\mu\nu\rho\sigma} p_{3\rho}p_{4\sigma} \varepsilon^*_\mu (p_3,\lambda) \varepsilon^*_\nu (p_4,\lambda^\prime),
\end{eqnarray}
where $s,s^\prime$ and $\lambda,\lambda^\prime$ are the spin and polarization labels, respectively. With this ampltiude, the leading order of thermally average annihilation cross section reads
\begin{eqnarray}
    \langle \sigma v\rangle = \log{(2)}^4\frac{\alpha_{\rm EM}^2 C_{ff}^2}{8\pi f_a^2 v_\varphi^2} \frac{m_f^4 m_{\chi}^2}{(m_a^2 - 4m_\chi^2)^2}.
\end{eqnarray}

We can see clearly from Fig.~\ref{fig: dm dm to photon photon} that the contribution of diphoton channel is sub leading compared to the $t\bar t$ channel. Therefore, this channel is not relevant to the DM relic density and the DM direct detection limit. The smallness is easily understood from $\alpha_{\rm EM}^2$ and $(m_f^2)/s$ suppressions.

\begin{figure}[t]
\centering
\includegraphics[width=0.6\textwidth]
{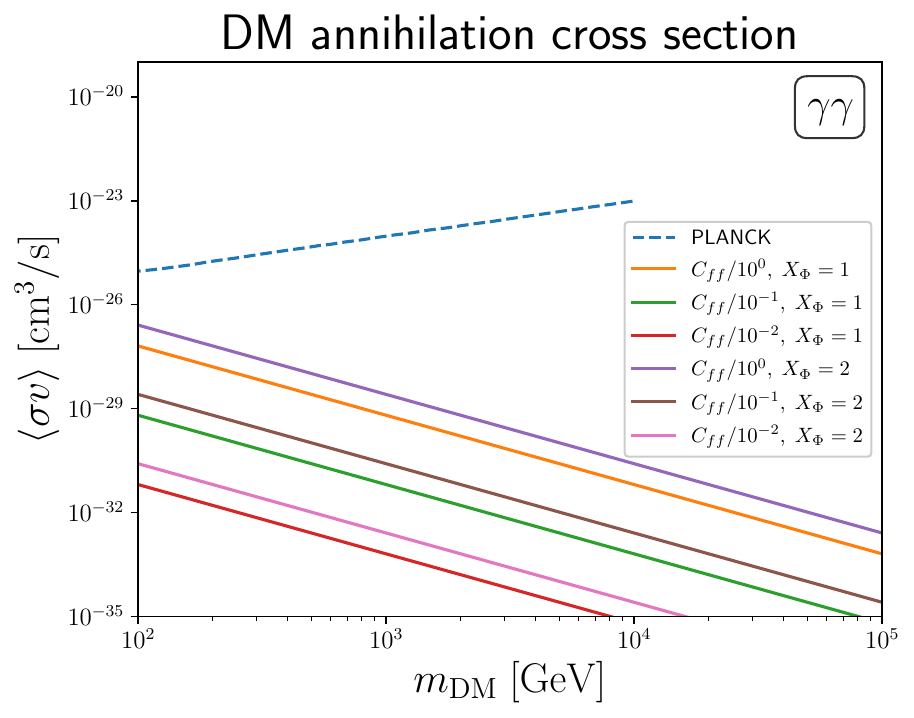}
\caption{The annihilation cross section of $\chi\bar\chi \to \gamma\gamma$.}
\label{fig: dm dm to photon photon}
\end{figure}

\bibliographystyle{jhep}
\bibliography{ref}
\end{document}